\documentclass[%
 reprint,
 amsmath,amssymb,
 aps,
 prc,
 floatfix,
 nofootinbib
]{revtex4-2}
\usepackage{hyperref}

\RequirePackage{tikz}
\RequirePackage{array}
\RequirePackage{mathtools}
\RequirePackage{multirow}

\renewcommand{\t}[1]{\text{#1}}

\begin{document}

\preprint{APS/123-QED}%
\title{Spurious Isospin Breaking in the In-medium Similarity Renormalization Group}
\author{A. Farren}%
\affiliation{%
School of Mathematics, Trinity College Dublin, Dublin,
Ireland
}%
\thanks{Present address: DAMTP, University of Cambridge, UK}
\author{S.~R.~Stroberg}%
\affiliation{
 Department of Physics and Astronomy,
 University of Notre Dame
 Notre Dame, IN, 46556, USA
}%
\email{rstroberg@nd.edu}
\date{\today}

\begin{abstract}

The in-medium similarity renormalization group is a promising ab initio approach for computing the isospin-related theoretical correction $\delta_C$ to superallowed beta decay rates, relevant for tests of CKM unitarity. An essential requirement of such corrections is a properly quantified theoretical uncertainty.
While isospin symmetry is explicitly broken by internucleon forces, approximations made in the solving the many-body problem can also spuriously break the symmetry, obscuring the desired signal. In this work, we use a toy model to identify the primary sources of spurious isospin symmetry breaking introduced by the IMSRG many-body framework, and provide mitigation strategies for application in large-scale calculations.

\end{abstract}


\maketitle

\section{Introduction}
Superallowed Fermi beta decays provide a stringent test of the Standard Model~\cite{TownerHardy2020}.
These tests require precise measurements, as well as precise calculations of Standard Model effects.
A current mild tension with unitarity of the Cabibbo-Kobayashi-Maskawa quark mixing matrix~\cite{Cabibbo:1963yz, Kobayashi:1973fv}, combined with recent progress in \textit{ab initio} nuclear theory has led to increased interest in revisiting these Standard Model corrections with a more rigorous uncertainty quantification~\cite{Cirigliano_2023, chakraborty2024workinggroup1summary}.
In the present work, we focus on the isospin-breaking correction $\delta_C$, defined~\cite{TownerHardy2020} as the deviation of the matrix element of the isospin raising operator from the isospin-conserving limit:
\begin{equation}
    |\langle \psi_f | T^{\pm} | \psi_i\rangle|^2 = |\left( \psi_f\right\vert T^\pm \left\vert \psi_i \right)|^2 ( 1-\delta_C).
\end{equation}
For the most extensively studied $T=1$ cases, the isospin-limit matrix element is $|\left( \psi_f\right\vert T^\pm \left\vert \psi_i \right)|^2=2$.
Deviations from this value are due primarily to the Coulomb interaction between protons, in addition to smaller effects such as the $u$-$d$ quark mass difference, and the mass difference between the charged and neutral pions~\cite{VanKolck1996,Cheung1980}.

The precisely-measured superallowed decays range from $A=10$ to $A=74$, and in all cases the initial or final nucleus (or both) is open shell, so these decays pose a challenge for \textit{ab initio} theory.
The no-core shell model~\cite{barrett2013,Caurier2002} and quantum Monte Carlo~\cite{Piarulli2026} can reach the lightest of these systems, but reaching beyond $A\sim 14$ will require a method that scales polynomially with the system size.
The in-medium similarity renormalization group (IMSRG)~\cite{Hergert2016}, and in particular the valence space formulation (VS-IMSRG)~\cite{Stroberg2019}, is a polynomially-scaling method capable of reaching all relevant systems.
However, as with other polynomially-scaling methods, quantifying uncertainties due to truncations in the many-body solution is non-trivial.
A first exploration of the $\delta_C$ correction within the IMSRG~\cite{Stroberg2021} found significant reference dependence, especially in lighter systems, as well as slow convergence with basis size.
As we will describe below, the reference dependence can be traced back to \emph{spurious} isospin symmetry breaking due to truncations in the IMSRG.
This spurious breaking directly competes with the authentic and physical signal of interest.

In this paper, we explore the sources and mechanisms of spurious symmetry breaking within the IMSRG, with the eventual goal of reducing these effects to levels needed for precise tests of the Standard Model.
While we focus on isospin symmetry breaking, the mechanisms we identify here also affect the breaking of other symmetries, for example of rotational symmetry in the treatment of deformed systems~\cite{Yuan2022}.
However, there is an important difference in that when \emph{exact} symmetries are spuriously broken by approximations, the symmetry may be restored by projection techniques~\cite{RingSchuck1980,Sheikh2021}.
This solution is not available in the case of isospin, where the symmetry is explicitly broken, and the degree of symmetry breaking is the quantity of interest.
One can develop strategies in which projection techniques are used to generate a symmetry-preserving basis, followed by a diagonalization of the symmetry-breaking part of the Hamiltonian~\cite{Satula2012}, and we briefly discuss that here.

Before investigating symmetry breaking within the IMSRG, we briefly review in section~\ref{sec:EOM} mechanisms of symmetry-breaking in wave function based approximation schemes, including Hartree-Fock, coupled cluster, and the equations of motion.
These can be compared with the IMSRG, in which operators are the central object.
A third class of methods are formulated in terms of approximations to the in-medium Green's function~\cite{Soma2020,Barbieri2016,Dickhoff2004,DickhoffMBT}.
For Green's function theories, there exists a framework for generating symmetry-conserving approximations, introduced by Baym and Kadanoff~\cite{Baym1961}, in which the self-energy is is obtained from a derivatives of a generating functional.
The random phase approximation (RPA), which we briefly discuss in section~\ref{sec:EOM}, is an example of such a conserving approximation~\cite{DickhoffMBT}.
However, we do not pursue the Baym-Kadanoff framework here.

In section~\ref{sec:IMSRG} we outline the relevant aspects of the IMSRG. In section~\ref{sec:spuriousbreaking} we use an isospin-conserving Hamiltonian to demonstrate spurious isospin breaking due to truncations made in the IMSRG, and we inspect the various mechanisms at play.
In section~\ref{sec:deltaC}, we look more specifically at the $\delta_C$ correction, including the effects of explicit isospin breaking terms in the Hamiltonian.
In the conclusion, we provide guidance for controlling spurious isospin breaking in large-scale calculations.

\section{Symmetry breaking in Hartree-Fock and Equations of Motion\label{sec:EOM}}

It has been long known that the Hartree-Fock ground state for $N\neq Z$ spuriously (spontaneously) breaks isospin symmetry~\cite{Brink1970,Engelbrecht1970,Lane1980}, even if the Hamiltonian is isospin symmetric.
Intuitively this is because, even with good isospin, the $pn$ interaction is stronger than the $pp$ or $nn$ interaction, so for an asymmetric system with $N\neq Z$, the protons and neutrons see a different mean field.
In this case, the Hartree-Fock ground state is not an eigenstate of $T^2$, but the Hamiltonian expressed in the Hartree-Fock basis still commutes with $T^2$.
Including correlations beyond the mean field then recovers the correct ground state isospin~\cite{Brink1970,Engelbrecht1970,Lane1980}.

In the equations of motion (EOM) framework~\cite{Rowe1968,Rowe2010,RingSchuck1980}, excited states $|\nu\rangle$ are expressed in terms of a raising operator acting on the ground state $|0\rangle$, while the ground state is annihilated by the corresponding lowering operators:
\begin{equation}
    |\nu\rangle = \chi_\nu^{\dagger}|0\rangle \quad,
    \qquad
    \chi_\nu |0\rangle = 0.
\end{equation}
The Schr\"odinger equation may be written as
\begin{equation} \label{eq:EOM}
    [H,\chi_\nu^\dagger] |0\rangle = \omega_\nu \chi_\nu^{\dagger}|0\rangle,
\end{equation}
where $\omega_\nu$ is the excitation energy of $|\nu\rangle$, relative to $|0\rangle$.

We consider an operator $Q$ which is the generator of a continuous symmetry of the Hamiltonian, which implies $[H,Q]=0$.
If $|0\rangle$ is not an eigenstate of $Q$, then $\chi_\nu^{\dagger}=Q$ is a solution of~\eqref{eq:EOM} with $\omega_\nu=0$, corresponding to a Nambu-Goldstone mode~\cite{Thouless1961}.
A relevant concrete example is the $x$ component of isospin $Q=T_x=\tfrac{1}{2}(T^++T^-)$; acting $T_x$ on a state with non-zero isospin connects to isobaric analogue states, which are degenerate in the isospin-symmetric limit. 

In general, it is not feasible to solve~\eqref{eq:EOM} exactly, and some approximation must be made.
Typically, one truncates the operator basis in which the $\chi_\nu^{\dagger}$ are expressed.
If $Q$ (or, more specifically, the action of $Q$ and $[H,Q]$ on $|0\rangle$) can be fully represented in the truncated operator basis, then the $\omega_\nu=0$ mode remains, and the symmetry is not broken~\cite{Rowe1968,RingSchuck1980,Rowe2010}.

In the Tamm-Dancoff approximation (TDA), the raising operators are constructed from one-particle-one-hole excitations
\begin{equation}\label{eq:chiTDA}
    \chi_\nu^{\dagger} = \sum_{ai} C^{\nu}_{ai}a^\dagger i
\qquad\text{(TDA)}.
\end{equation}
The Hartree-Fock ground state (or any other product state defining the hole orbits $i$) is annihilated by the corresponding $\chi_\nu$.
If the Hermitian symmetry operator $Q$ has particle-hole excitation matrix elements of the form~\eqref{eq:chiTDA}, then it must also have de-excitation matrix elements.
These contract with $H$ to make a 1p1h contribution to $[H,Q]$ which cannot be captured by the truncated $\chi_\nu^{\dagger}$, and so the TDA will break symmetry.
As a concrete example, $T_x$ will have particle-hole matrix elements for a reference with $N\neq Z$, so the TDA will spuriously break isospin symmetry.

In the random phase approximation (RPA), the raising operator $\chi_\nu^{\dagger}$ contains both excitation and de-excitation pieces:
\begin{equation}\label{eq:chiRPA}
    \chi_\nu^{\dagger} = \sum_{ai}X^{\nu}_{ai}a^{\dagger} i - Y^{\nu}_{ai} i^{\dagger}a
   \qquad\text{(RPA)}.
\end{equation}
This implies that the ground state, which is annihilated by $\chi_\nu$ is not a simple product state, i.e. it is correlated.
If $Q$ is a one-body operator, then it can be represented in the form~\eqref{eq:chiRPA} and the RPA preserves the corresponding symmetry~\cite{RingSchuck1980, Engelbrecht1970,Lane1980,Rowe2010}, though we note some subtleties regarding the stability of the Hartree-Fock ground state~\cite{Lane1980}.
Note that this does not guarantee that \emph{explicit} symmetry-breaking effects will be captured accurately by the RPA.
In addition, improving the RPA by including excitations beyond 1p1h, e.g. via second RPA~\cite{Sawicki1962}, will in general lead to symmetry breaking~\cite{Rowe2010}.
These arguments extend also to open shell systems treated with the quasiparticle RPA~\cite{RingSchuck1980,Simkovic2013}.

\newcommand{\CCT}{\ensuremath{\mathcal{T}}}
In the coupled cluster (CC) version of EOM~\cite{Stanton1993,ShavittBartlett,Hagen2014,Ekstrom2014}, the ground state $|0\rangle$ is taken to be the coupled cluster ground state\footnote{To avoid confusion, we use the symbol $\CCT$ for the cluster operator, and reserve $T$ for isospin.}
$e^{\CCT}|{\rm HF}\rangle$.
Obtaining operators $\chi_\nu$ which annihilate the CC ground state is non-trivial in general.
Instead,~\eqref{eq:EOM} is solved using the similarity-transformed Hamiltonian $\bar{H}=e^{-\CCT}He^{\CCT}$, and similarity transformed (de-)excitation operators\footnote{Because $\CCT^{\dagger}\neq -\CCT$, we have $\bar{\chi}_\nu^{\dagger}\neq (\bar{\chi}_{\nu})^{\dagger}$.}
\begin{equation}
    \bar{\chi}_\nu^{\dagger} = e^{-\CCT} \chi_\nu^{\dagger} e^{\CCT}
    \quad , \qquad
    \bar{\chi}_\nu = e^{-\CCT} \chi_\nu e^{\CCT}
\end{equation}
that satisfy $\bar{\chi}_\nu|{\rm HF}\rangle\!=\!0$.
The similarity-transformed operator $\bar{Q}=e^{-\CCT}Qe^{\CCT}$ commutes with $\bar{H}$, and so the corresponding symmetry will be conserved if the basis used for $\bar{\chi}_{\nu}^{\dagger}$ is sufficient to represent $\bar{Q}$ and $[\bar{H},\bar{Q}]$.
In EOM-CC, $\bar{\chi}_\nu^{\dagger}$ is taken to be a pure excitation operator.
This does not cause the issue discussed above in the TDA case because---by design---$\bar{H}$ does not have any 1p1h or 2p2h excitation components so any de-excitation parts of $\chi_\nu^{\dagger}$ cannot contribute to~\eqref{eq:EOM}.
However, when the cluster operator $\CCT$ is truncated at the two-body level, corresponding to coupled cluster with singles and doubles, and $Q$ has non-zero particle-hole matrix elements, then $\bar{Q}$ will have a three-body component arising from $[[Q,\CCT],\CCT]$, as illustrated in Fig.~\ref{fig:CCdiagram}.
This means that in order to fully represent $\bar{Q}$, the EOM excitation operator $\chi_\nu^{\dagger}$ should include 3p3h excitations, while representing $[\bar{H},\bar{Q}]$ requires 4p4h excitations.
Typically the EOM excitation operator is truncated at the same level as the cluster operator, so we expect EOM-CC to spuriously break symmetry for an asymmetric reference.

As a side note, a classic example of a zero mode in finite nuclei is the translation generated by the center of mass momentum operator~\cite{Thouless1961,RingSchuck1980,Rowe2010}.
The above discussion suggests that center-of-mass motion is not cleanly separated out in the EOM-CC as it is in the RPA, so that the $J^\pi=1^-$ spectrum will have some contamination even in the limit of a complete single-particle basis.
However, in practical calculations this appears to be a minor issue that can be mitigated with proper analysis~\cite{Hagen2009,Jansen2013}.

While techniques have been developed to restore broken symmetries via projection in coupled cluster theory~\cite{Duguet2015,Qiu2017,Hagen2022}, as discussed in the introduction such methods are not directly applicable in the presence of explicit symmetry breaking.

\begin{figure}
    \centering
    \includegraphics[width=0.75\linewidth]{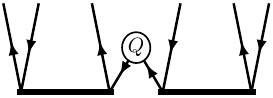}
    \caption{Diagram representing contribution to $\bar{Q}$ due to $[[Q,\CCT],\CCT]$ in coupled cluster, in the case where $Q$ has non-zero particle-hole de-excitation matrix elements.}
    \label{fig:CCdiagram}
\end{figure}

\section{IMSRG\label{sec:IMSRG}}
The methods discussed in section~\ref{sec:EOM} are schemes for approximating the wave function.
In contrast, the IMSRG focuses primarily on operators, and truncations are made to the space of operators.
Consequently, the specific mechanisms by which symmetries are broken will in general be different from those discussed in section~\ref{sec:EOM}.
An EOM extension of the IMSRG exists~\cite{Parzuchowski2017x,Parzuchowski2017,Porro2026}, with some similarities to the EOM-CC, but an attractive feature of the IMSRG is the convenient formulation in a valence space, which enables straightforward treatment of the open-shell systems of interest for superallowed Fermi decays.
The details of the IMSRG approach are presented in several review articles~\cite{Hergert2016,Hergert2016PhysRep,Stroberg2019}.
Here we briefly summarize the relevant features.

To simplify the many-body calculation, we partition the Hamiltonian $H=H^{\rm d}+H^{\rm od}$ into a block-diagonal piece $H^{\rm d}$ and an off-diagonal piece $H^{\rm od}$.
We then construct a unitary transformation $U$ to block-diagonalize the Hamiltonian, and subsequently solve the problem in the reduced space.
The unitary transformation is constructed through a flow equation, so that 
\begin{equation} \label{eq:dUds}
    \frac{dU}{ds} = \eta(s) U(s)
\end{equation}
where the anti-hermitian operator $\eta=-\eta^\dagger$ is the generator of the transformation.
With a suitable choice of generator, the transformed Hamiltonian
\begin{equation}
    \label{eq:srg}
    H(s)\equiv U(s) H(0)U^\dagger(s)
\end{equation}
can be driven to block-diagonal form, i.e. $H^{\rm od}(s)\to 0$
as $s\to \infty$.
Combining \eqref{eq:dUds} and \eqref{eq:srg}, we obtain a flow equation for $H(s)$ in terms of a commutator with $\eta$:
\begin{equation} \label{eq:dHds}
    \frac{dH}{ds} = [\eta(s),H(s)].
\end{equation}
Other operators can be consistently transformed $\mathcal{O}=U(s)\mathcal{O}(0)U^{\dagger}(s)$.

In the Magnus formulation~\cite{Magnus}, the unitary transformation is parameterized as the exponential of an antihermitian operator $\Omega$, i.e. $U=e^{\Omega}$.
Then~\eqref{eq:dUds} corresponds to a flow equation
\begin{equation}\label{eq:magflow}
    \frac{d\Omega}{ds} = \sum_{k=0}^{\infty} \frac{B_k}{k!}[\Omega(s),\eta(s)]^{(k)}
\end{equation}
where $B_k$ is the $k$th Bernoulli number and $[\Omega,\eta]^{(k)}$ indicates a $k$-fold nested commutator, e.g. $[\Omega,\eta]^{(2)}\equiv [\Omega,[\Omega,\eta]]$.
The calculations presented here all use the Magnus formulation, but we have verified that our conclusions hold for the flow equation formulation~\eqref{eq:dHds}.

All operators are represented in a basis of Fock-space operators.
Specifically, we write the Hamiltonian as
\begin{equation}\label{eq:HNO}
\begin{aligned}
    H =& E_0
    + \sum_{ij} f_{ij}\{ i^{\dagger} j\} 
    + \frac{1}{(2!)^2} \sum_{ijkl}\Gamma_{ijkl} \{i^{\dagger}j^{\dagger}lk \} \\
    &+ \frac{1}{(3!)^2} \sum_{ijklmn} W_{ijklmn} \{ i^{\dagger}j^{\dagger}k^{\dagger}nml\}
    +\ldots
    \end{aligned}
\end{equation}
where $i^{\dagger}$ creates a nucleon in single-particle state $i$, and the braces $\{\}$ indicate normal ordering with respect to some reference state $|\Phi_0\rangle$.
If all possible strings of creation/annihilation operators are retained, we have a complete operator basis, regardless of the choice of $|\Phi_0\rangle$.
In practical calculations, we truncate the operator basis, typically retaining up to two-body operators in the IMSRG(2) approximation.
Due to this truncation, the computed observables depend on the choice of reference.

We work in a single-particle basis of spherical harmonic oscillator eigenstates and truncate this basis by limiting the quantum number $(2n+\ell)\leq e_{\rm max}$.
Unless otherwise indicated, we then transform all operators to the Hartree-Fock basis before performing the IMSRG calculation.

In this work, we perform IMSRG calculations up to the IMSRG(3) level for both the flowing Hamiltonian and for the consistent transformation of the $T^{\pm}$ operator.
The latter required additional code implementation beyond what was described in previous work~\cite{Heinz2021,stroberg2024imsrgflowing3body,Heinz2025}.
Artificial intelligence tools (Claude Opus 5.5) were used to identify bugs and suggest optimizations in the new code.
All suggested changes were manually verified and validated with deterministic unit tests.
The code and its revision history can be found at~\cite{IMSRG++}.
The results presented here can be reproduced with version {devel\_e50f772}.

\subsection{Single-reference and valence-space IMSRG}
In the simplest implementation, which we denote single-reference IMSRG, the IMSRG flow is used to decouple the reference $|\Phi_0\rangle$ from all excitations.
In this case, $H^{\rm od}$ consists of all terms in~\eqref{eq:HNO} which, when acting on $|\Phi_0\rangle$, would lead to configurations other than $|\Phi_0\rangle$, e.g. $\Gamma_{pp'hh'}$, where $p,p'$ are particle indices and $h,h'$ are hole indices.

\begin{figure}[t]
    \centering
    \includegraphics[width=\columnwidth]{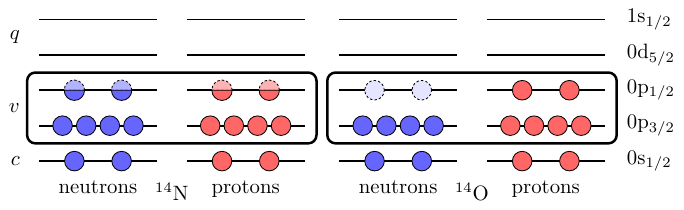}
    \caption{$p$-shell valence space configuration of $^{14}$N and $^{14}$O, where orbitals are core ($c$), valence ($v$) or excluded ($q$). Nucleons in the 0p$_{1/2}$ shell of $^{14}$N are half-filled in the ensemble reference.
    }
    \label{fig:O14N14valence}
\end{figure}

In the valence space IMSRG (VS-IMSRG) \cite{Bogner2014,Stroberg2017valence} we partition the single-particle orbits into core, valence and excluded orbits, labeled $c,v,q$ respectively in Fig.~\ref{fig:O14N14valence}.
The sub-block to be decoupled consists of all configurations in which all core orbits are occupied and all excluded orbits are unoccupied.
Valence orbits can be either occupied or unoccupied.
The Hamiltonian is then diagonalized in this subspace.
In this case, $H^{\rm od}$ consists of all terms which, when acting on a valence configuration would lead to a configuration with an unoccupied core orbit or an occupied excluded orbit, e.g. $\Gamma_{qv'vc}$.
More details can be found in the reviews~\cite{Hergert2016,Stroberg2019}.

\subsection{Choice of generator}
We have the freedom to chose $\eta(s)$ as long as it drives $H^\t{od}(s)\to0$ along the flow. Here we introduce three popular choices of generator.
The White generator~\cite{White2002} is defined schematically as
\begin{equation}\label{eq:white1}
    \eta^{\rm Wh}(s) \equiv \frac{H^\t{od}(s)}{\Delta(s)} 
\end{equation}
where we introduce an energy denominator $\Delta$.
The \textit{arctangent generator}~\cite{White2002} is 
\begin{equation}
\label{eq:atan}
    \eta^{\rm atan}(s) \equiv \frac{1}{2}\mathrm{atan}(2\frac{H^\t{od}(s)}{\Delta(s)})
\end{equation}
while the \textit{imaginary time generator}~\cite{Hergert2016PhysRep} is 
\begin{equation}\label{eq:IT}
    \eta^{\rm IT}(s) \equiv \mathrm{sgn}{\Delta(s)} H^\t{od}(s).
\end{equation}
More detailed expressions can be found in Ref.~\cite{Hergert2016PhysRep}.
The solution to the flow equation \eqref{eq:dHds} for all three choices of $\eta$ leads to suppressed $H^{\rm od}$ \cite{Hergert2016PhysRep}. In IMSRG calculations, we primarily use denominators which are partitioned based on the Epstein-Nesbet or M{\o}ller-Plesset prescription.
The Epstein-Nesbet partitioning of an energy denominator $\Delta_{ab...ij...}$ is defined as follows, up to a particle rank of 2, 
\begin{equation}
    \label{eq:DeltaEN1}
    \begin{aligned}
        \Delta_{ai}&=\langle \Phi_i^a| H | \Phi_i^a\rangle-\langle \Phi | H | \Phi\rangle\\
    &=f_{aa}-f_{ii}+\Gamma_{aiai},
    \end{aligned}
\end{equation}
and
\begin{equation}
    \label{eq:DeltaEN2}
    \begin{aligned}
        \Delta_{abij}&=\langle \Phi_{ij}^{ab}| H| \Phi_{ij}^{ab}\rangle-\langle \Phi|H|\Phi\rangle\\
        &=f_{aa}+f_{bb}+\Gamma_{abab}-f_{ii}-f_{jj}-\Gamma_{ijij} \\ 
        & \qquad -\Gamma_{aiai}-\Gamma_{bjbj}-\Gamma_{ajaj}-\Gamma_{bibi}.
    \end{aligned}
\end{equation}
The generator is manifestly anti-hermitian since $\Delta_{ia}=-\Delta_{ai}$, $\Delta_{ijab}=-\Delta_{abij}$.
As we discuss later, when taking advantage of the rotational symmetry of the Hamiltonian, we couple two-body matrix elements to good total angular momentum $J$, denoted $\Gamma_{ijkl}^{J}$, but the energy denominators are formed with the monopole component.
The M{\o}ller-Plesset partitioning of an energy denominator is the same as Epstein-Nesbet, but this time omitting the two-body matrix elements $\Gamma_{pqrs}$.

\section{Spurious isospin symmetry breaking in the IMSRG\label{sec:spuriousbreaking}}
To illustrate the spurious symmetry-breaking in the IMSRG, we begin with an isospin-conserving Hamiltonian and compute the ground-state expectation value of the isospin-squared operator $T^2$ with the IMSRG.
In this case, any deviation from the exact answer indicates the extent to which truncations made in solving the IMSRG flow equations break isospin symmetry.
We focus on the decay $^{14}{\rm O}\to^{14}{\rm N}$, which has the advantage that it can be treated in the single-reference (supplemented with the equations-of-motion method~\cite{Rowe1968,Parzuchowski2017}) or valence-space formulations.
The choice of isospin-conserving Hamiltonian we make is 
\begin{equation}
    H = T_{\rm rel} +V_{\rm iso}
\end{equation}
where $T_{\rm rel}$ is the relative kinetic energy, and $V_{\rm iso}$ is an isospin-conserving NN potential.
Here we use a modified version of the $\Delta$NNLO$_{GO}$ potential of Ref.~\cite{Jiang2020deltaGO}, without the Coulomb interaction.
This interaction includes isospin-breaking strong interactions, as well as an isospin-conserving 3N force.
For convenience, we normal-order with respect to an $^{16}$O reference state built from a single Slater determinant of harmonic oscillator wave functions, and discard the residual 3N force.
Then we isospin-average the resulting NN potential to yield an isospin-conserving potential $V_{\rm iso}$, which we use for our subsequent investigation.

\begin{figure*}
    \centering
    \includegraphics[width=0.95\textwidth]{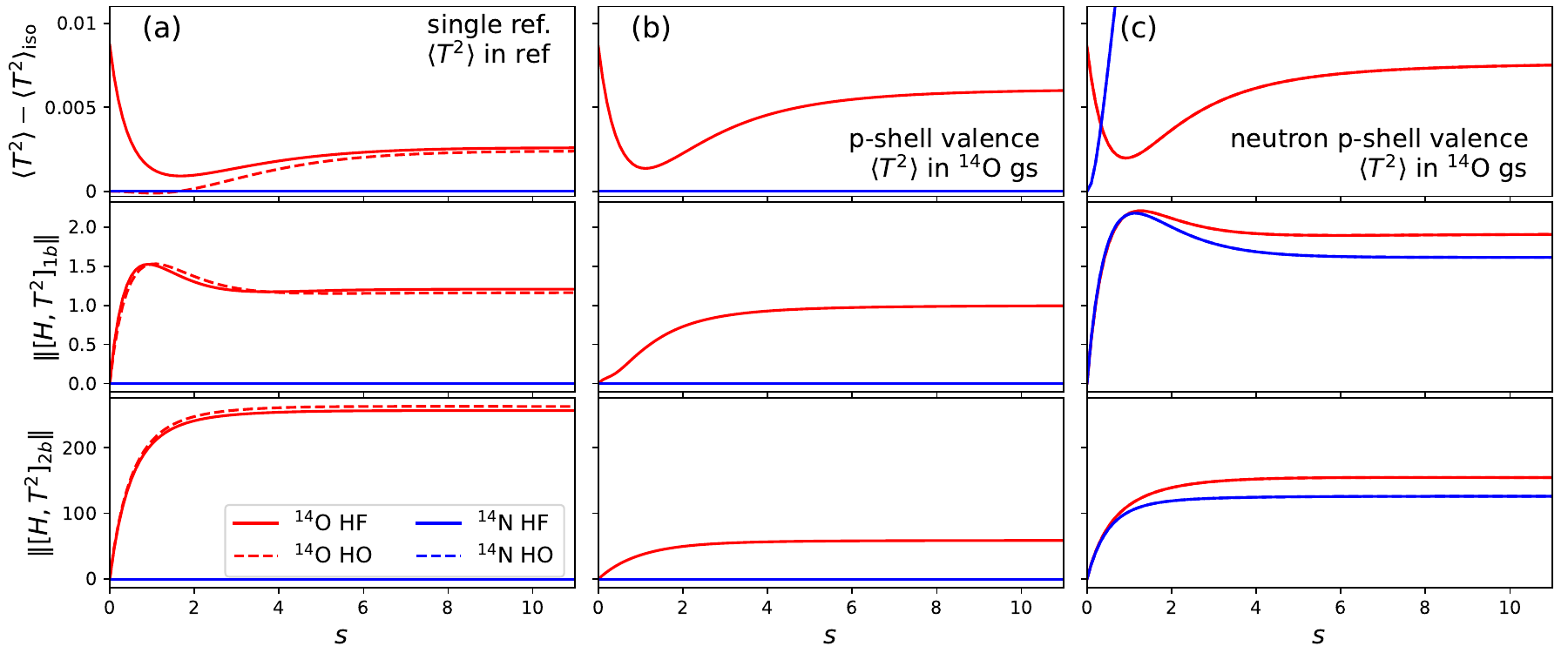}

    \caption{Isospin symmetry breaking along the IMSRG(2) flow for $^{14}$O and $^{14}$N (see text for details). The top row shows the expectation value of $T^2$, relative to the exact value. The middle and bottom rows show the norm of the one-body and two-body pieces, respectively, of the flowing commutator $[H,T^2]$.
    In column (a) we use a single-reference decoupling. In column (b) we decouple a $p$-shell valence space, and compute $\langle T^2\rangle$ in the $^{14}$O ground state. In column (c) we employ a valence space of neutrons only.
    }
    \label{fig:O14N14T2HT}
\end{figure*}

We take as the initial reference state a single Slater determinant $|\Phi_0\rangle$ formed by occupying the lowest $Z$ proton states and $N$ neutron states, in either the harmonic oscillator or Hartree-Fock basis.
We integrate the flow equation~\eqref{eq:dHds} or~\eqref{eq:magflow} to decouple $|\Phi_0\rangle$ from excitations and consistently evolve the $T^2$ operator.
For $s\to\infty$, the zero-body piece of $T^2(s)$ gives the ground-state expectation value.
For an $^{14}$O reference, with $Z=8$, $N=6$, an isospin-conserving interaction should yield $\langle T^2\rangle=2$.
We also compute the commutator of the flowing operators $[H(s),T^2(s)]$.
The operators commute at $s=0$, and to the extent that the IMSRG transformation is unitary, these operators should commute throughout the flow.
For comparison, we repeat the calculation with a $^{14}$N reference, which has $N\!=\!Z\!=\!7$.
Since $^{14}$N is open shell, we use an ensemble reference in which each of the four $0p_{1/2}$ orbits is occupied with probability 0.5.
This ensemble is not an eigenstate of particle number, but it can nevertheless be used for normal ordering~\cite{Kutzelnigg1997,Stroberg2017valence}. The zero-body part of the $T^2$ operator corresponds to a trace over the ensemble, which in the case of $^{14}$N gives $\langle T^2\rangle= 3/4$ if isospin is conserved.
The results of these calculations are shown in Fig.~\ref{fig:O14N14T2HT}(a).

We observe that for the $N\!\neq \!Z$ $^{14}$O reference, isospin symmetry is broken in the ground state, and at the level of operators.
For the $^{14}$N reference, isospin conservation is maintained at the level of the numerical precision of the calculation.
This is in line with the expectations set in section \ref{sec:EOM}, whereby we concluded that a reference with $N\!\neq\!Z$ is generally expected to spuriously break isospin in truncated computational schemes.
In this section, we investigate how this happens in practical IMSRG implementations and suggest remedies.

Before proceeding, we point out that the choice of a harmonic oscillator or Hartree-Fock single-particle basis appears to have very little effect on the final result. To see this, notice the following. In Fig.~\ref{fig:O14N14T2HT}(a) we see at $s=0$ that the Hartree-Fock reference breaks isospin in the ground state, while the harmonic oscillator reference preserves it.
The IMSRG evolution initially brings the Hartree-Fock result closer to the exact value, but as $s\to\infty$ the symmetry remains broken.
The near-independence of this symmetry breaking on the single-particle basis suggests that the Hartree-Fock step is not primarily to blame, and we should investigate the IMSRG part of the calculation.

In the IMSRG, the essential approximation is the truncation of the basis of operators during the flow, typically based on particle rank.
If the preservation of a symmetry depends on a cancellation between terms which are impacted differently by this truncation, the symmetry may be broken.
An analogous situation arises in many-body perturbation theory (MBPT): if the preservation of the symmetry requires cancellations between different perturbative orders, then truncating to a given order will tend to break the symmetry.

In both cases, we can frame the issue in terms of a partitioning of the Hamiltonian in such a way that the individual partitions do not preserve the symmetry.
In MBPT, the Hamiltonian is partitioned as $H=H_0+V$, where $H_0$ is exactly solvable; working in the eigenbasis of $H_0$ we truncate based on powers of the perturbation $V$. (When using the Hartree-Fock basis, this perturbation is the residual interaction left over after accounting for the mean field.)
In the IMSRG, we partition the Hamiltonian into ``diagonal'' and ``off-diagonal'' pieces $H=H^{\rm d}+H^{\rm od}$, where $H^{\rm od}$ should be suppressed by the IMSRG flow.
The flowing operators are then truncated based on particle rank.
Since we use normal-ordered operators, both of these partitions in general depend on the choice of reference $|\Phi_0\rangle$.
In addition, the IMSRG will be sensitive to a symmetry-breaking generator $[\eta,T^2]\neq0$. We now explain these dependencies.

\begin{figure*}
    \centering
    \includegraphics[width=0.95\textwidth]{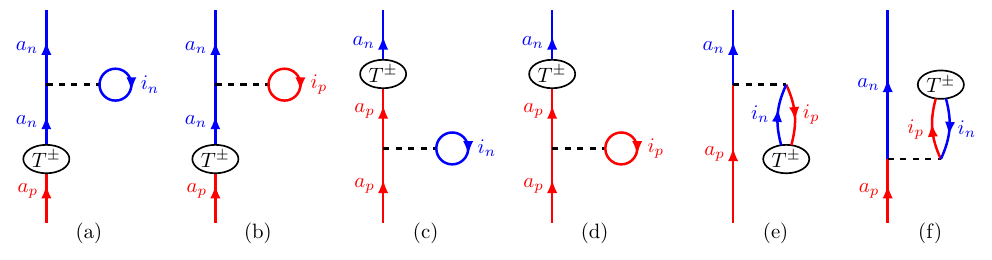}
    \caption{\label{fig:CommHT1b}Antisymmetrized Goldstone diagrams representing contributions to the commutator $[H,T^{\pm}]$. If both $i_p$ and $i_n$ are occupied, diagram (a) cancels with diagram (d), and (b) cancels with (c). If $i_n$ is not occupied, then diagrams (a) and (c) do not contribute, but (e) does. Diagram (f) also contributes for an ensemble reference in which $i_n$ and $i_p$ are fractionally filled.}
\end{figure*}

\subsection{Dependence on reference state}
We first illustrate the impact of reference choice with a simple example of the isospin raising/lowering operator $T^\pm$, which commutes with an isospin-symmetric Hamiltonian.
We consider two spin-spatial orbits $a$ and $i$, for both protons and neutrons, leading to four possible single-particle states, such that $T^+|a_n\rangle = |a_p\rangle$, and $T^+|i_n\rangle = |i_p\rangle$.
First, we take $i_p,i_n$ to be occupied in the reference, while $a_p,a_n$ are unoccupied.
In the Moller-Plesset partitioning $H_0$ consists of the kinetic energy (which conserves isospin to the extent that the proton and neutron masses are equal) plus the normal-ordered one-body part of the potential.
Focusing on the matrix element of the commutator $[H,T^{\pm}]$ between one-particle states, the four antisymmetrized Goldstone diagrams which contribute are shown in Fig.~\ref{fig:CommHT1b}(a) through (d), corresponding to
\begin{equation}\label{eq:diagramsabcd}
\begin{aligned}
\bigl\langle a_n\bigl|\, [H,T^{\pm}] \, \bigr| a_p \bigr\rangle  = ({\rm a})+({\rm b}) - ({\rm c}) -({\rm d}),
\end{aligned}
\end{equation}
while diagrams (e) and (f) do not contribute.
Explicit expressions for the diagrams can be found in Appendix~\ref{app:diagrams}.
The commutator vanishes for an isospin conserving potential $V$.
Specifically, in the isospin limit
$(\mathrm{a})=(\mathrm{d})$,
$(\mathrm{b})=(\mathrm{c})$, and 
$(\mathrm{e})=(\mathrm{f})$.
More generally, we can specify the occupations $n_{i_p},n_{i_n}$, where $0\leq n_{i_p},n_{i_n} \leq 1$, in which case,~\eqref{eq:diagramsabcd} becomes
\begin{equation}\label{eq:abcdWithOcc}
\begin{aligned}
    \bigl\langle a_n \bigl| [H,T^\pm] \bigr| a_p \bigr\rangle
    &=n_{i_n} (\mathrm{a})
    + n_{i_p} (\mathrm{b})
    - n_{i_n} (\mathrm{c}) 
    - n_{i_p} (\mathrm{d})    \\
    +& n_{i_p}(1-n_{i_n})(\mathrm{e})
    - n_{i_n}(1-n_{i_p})(\mathrm{f}) .
    \end{aligned}
\end{equation}

If we use an asymmetric reference with $n_{i_p}=1$ and $n_{i_n}=0$, then diagrams (a) and (c) no longer contribute.
Diagrams (b) and (d) are not equal, since (b) acts in the $T\!=\!0$ and $T\!=\!1$ channels, while (d) only acts in $T\!=\!1$, and so they do not combine to give zero.
However, now diagram (e) contributes.
It acts in the $T\!=\!0$ channel and exactly cancels the difference $({\rm b})\!-\!({\rm d})$, preserving the isospin symmetry of the Hamiltonian.
Diagram (e) is first-order in the perturbing potential $V$, consisting of the two-body part of the normal-ordered interaction contracting with the one-body $T^{\pm}$ operator to yield a one-body operator. In a perturbative approach, conservation of isospin requires cancellations between terms of different order,
so that a truncation to a given order in perturbation theory will spuriously break isospin.
Likewise, if we were to truncate all operators at the normal ordered one-body level, diagram~\ref{fig:CommHT1b}(e) would be omitted and isopin would not be conserved.
Similar effects occur at higher particle rank, as would be induced during the IMSRG evolution, so discarding flowing three-body operators also leads to spurious isospin breaking.

As mentioned above, the $^{14}$N reference we use is an ensemble, for which all $0p_{1/2}$ orbits have occupation $n=\tfrac{1}{2}$.
From~\eqref{eq:abcdWithOcc}, which still holds for an ensemble reference, we see that for $n_{i_n}=n_{i_p}=\tfrac{1}{2}$, the commutator involves  cancellations between terms of the same particle rank (and the same perturbative order).
This holds regardless of how accurately the ensemble approximates the true eigenstate.

\subsection{ Impact of a symmetry-breaking generator }

Even with a symmetric reference state, a unitary transformation can reshuffle operators so that symmetry preservation again depends on cancellations between different particle ranks.
To illustrate this, we consider the IMSRG evolution of the commutator\footnote{We consider $T^2$ rather than $T^\pm$ because $T^2$ has a two-body piece, which better illustrates the effect of intermediate three-body operators. Similar examples can readily be made for $[H,T^{\pm}]$.} $[H,T^2]$.
This commutator is itself an operator that obeys the flow equation
\begin{equation}\label{eq:flowHT}
\begin{aligned}
    \frac{d}{ds}[H,T^2] &= [\eta,[H,T^2]]  \\
    &= [[\eta,H],T^2] + [H,[\eta,T^2]],
\end{aligned}
\end{equation}
where all operators are evaluated at flow parameter $s$.
If the initial Hamiltonian $H(0)$ commutes with $T^2$, then the first equality in \eqref{eq:flowHT} shows that they should commute for all $s$, assuming no truncation is made.
However, what is actually evaluated in the calculation corresponds to the second equality in~\eqref{eq:flowHT}; the $H$ and $T^2$ operators are separately evolved, and then we take the commutator of the evolved operators.
If the generator $\eta$ conserves isospin, then $\frac{d}{ds}T^2=[\eta,T^2]=0$, and the two terms in the second line of~\eqref{eq:flowHT} must therefore vanish independently.

Expanding out the double commutators in~\eqref{eq:flowHT} gives
\begin{equation}\label{eq:HdsTds eight terms}
    \begin{aligned}
        \frac{d}{ds}[H,T^2]&=
        (\eta H)T^2 -(H\eta)T^2-T^2(\eta H)+T^2(H\eta) \\
         & +H(\eta T^2)-H(T^2\eta)-(\eta T^2)H+(T^2\eta)H,
    \end{aligned}
\end{equation}
where the parentheses indicate which operator contraction is to be evaluated first. If $[\eta,T^2]=0$ then the second line of~\eqref{eq:HdsTds eight terms} vanishes. If in addition $[H,T^2]=0$, then the first and third terms in the first line of~\eqref{eq:HdsTds eight terms} should cancel, as should the second and fourth.
The contraction between $\eta$ and $H$ is performed first in both cases, and the terms with intermediate three-body operators should cancel each other.
Making the IMSRG(2) approximation discards pairs of terms which cancel (provided normal ordering is performed with respect to a symmetric reference, e.g. $N=Z$ or a symmetric ensemble reference) and so no spurious isospin breaking is induced.

\begin{figure}
    \centering
    \includegraphics[width=0.98\linewidth]{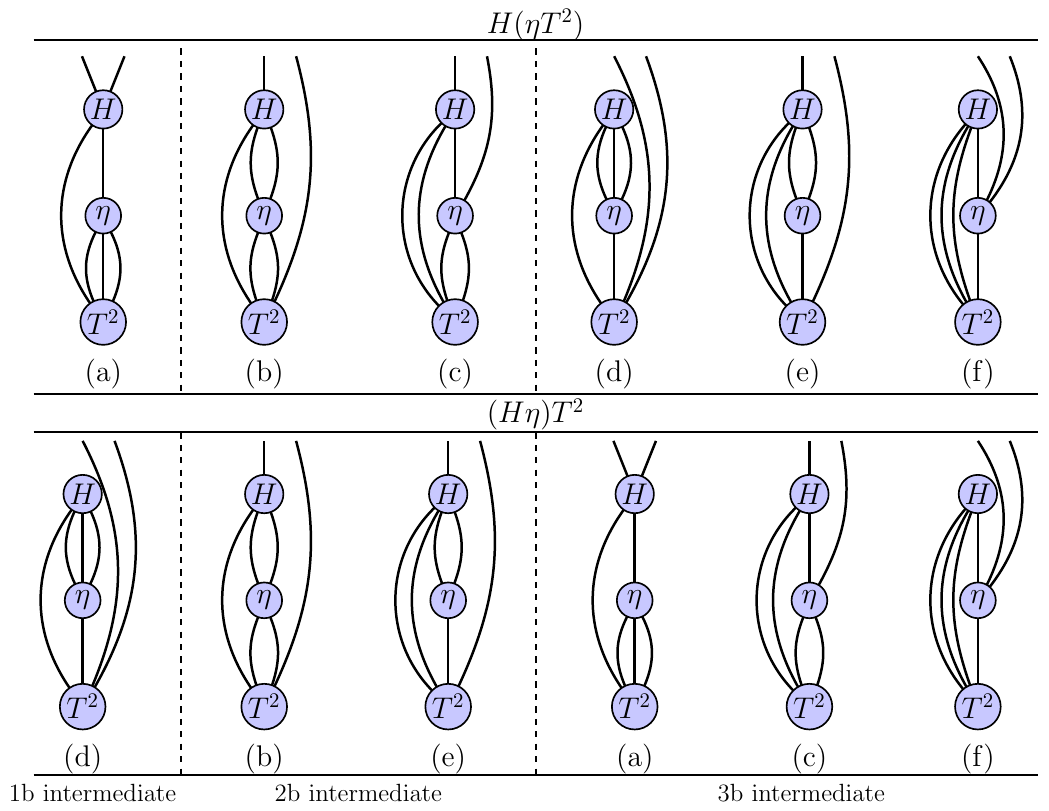}
        \caption{Hugenholtz skeleton diagrams~\cite{ShavittBartlett} with two-body parts of operators contributing to the one-body matrix elements of $\tfrac{d}{ds}[H,T^2]$ according to the flow equation \protect\eqref{eq:flowHT}. The top row should cancel exactly with the bottom, but isospin symmetry is broken in the IMSRG(2) approximation since some cancelling pairs have different intermediate particle rank.}
    \label{fig:HetaT2 diagrams}
\end{figure}

On the other hand, if $[\eta,T^2]\neq 0$, then we require cancellation between the two rows in~\eqref{eq:HdsTds eight terms}.
When we discard three-body operators induced by the commutator, the cancellation can be spoiled.
If $[H,T^2]=0$, then in the absence of truncation each term in the first row of~\eqref{eq:HdsTds eight terms} should cancel against a corresponding term in the second row.
For example, the $H(\eta T^2)$ term should cancel against $(H\eta)T^2$.
The one-body pieces of these terms are illustrated diagrammatically in Fig.~\ref{fig:HetaT2 diagrams}, where we group diagrams according to the particle rank of the intermediate operator indicated by parentheses.
We see that the diagram labeled (a) in Fig.~\ref{fig:HetaT2 diagrams} will be included in the IMSRG(2) approximation of $H(\eta T^2)$, but \emph{not} in $(H\eta)T^2$, where it involves a three-body intermediate; the IMSRG(2) approximation spoils the cancellation. 

The straightforward solution would be to include flowing three-body operators, performing an IMSRG(3) calculation.
While this is not feasible for large-scale calculations, medium-scale calculations can be performed with the IMSRG(3N7) approximation~\cite{Heinz2021,stroberg2024imsrgflowing3body,Heinz2025}, and large-scale calculations can be performed with the IMSRG(3f$_2$) approximation~\cite{he2024factorizedapproximationimsrg3}.
However, relaxing the particle-rank truncation fixes the issue for just one numerical flow step\footnote{In fact since the intermediate 3b diagrams in Fig.~\ref{fig:HetaT2 diagrams} are of the type $[2,2]\to3$, for $N\!\neq\!Z$ using IMSRG(3N7) is enough at one step to ensure $\frac{d}{ds}[H,T^2]=0$ while for $N=Z$ using IMSRG(2) is sufficient, as discussed following \eqref{eq:HdsTds eight terms}.}.
In the next step, there will be cancellations involving intermediate four-body operators, and so on.
We may still expect---or at least hope---that the importance of these terms will be small, so that going from IMSRG(2) to IMSRG(3) systematically reduces the magnitude of the spurious breaking.
This is explored in section~\ref{sec:deltaC}.

Having established the importance of an isospin-symmetric generator when performing the IMSRG truncation, we now identify several features which may lead to $[\eta,T^2]\neq 0$.

\subsubsection{ Partitioning of \texorpdfstring{$H=H^{\rm d}+H^{\rm od}$}{H} }

If the block-diagonal partitioning of the Hamiltonian is performed such that  $[H^{\rm od},T^2]\neq 0$, then in general we have $[\eta,T^2] \neq 0$ which, when combined with a truncation of the operator basis, can lead to spurious isospin-breaking.
In fact, this partitioning is the most significant source of isospin-breaking in the generator.

This is illustrated by the single-reference calculation of the $^{14}$O ground state shown in Fig~\ref{fig:O14N14T2HT}(a).
The reference $|\Phi_0\rangle$ consists of the proton orbits $\{\pi 0s_{1/2},\pi 0p_{3/2}, \pi 0p_{1/2}\}$ and the neutron orbits $\{\nu 0s_{1/2},\nu 0p_{3/2} \}$.
Excitations $\nu 0p_{3/2}\to \nu 0p_{1/2}$ connect the reference to excited configurations, and are included in $H^{\rm od}$, while $\pi 0p_{3/2}\to \pi 0p_{1/2}$ does not correspond to an excitation of the reference.
For standard choices of generator, this leads to $[\eta,T^2]\neq 0$ and consequently, as described in the previous section, involves cancellations which are spoiled when we truncate based on particle rank.

One way to mitigate this source of isospin breaking is to use a valence space which is (approximately) isospin symmetric, such as the $0p$ shell for both protons and neutrons.
In this case, the definition of off-diagonal is approximately the same for protons and neutrons, so that $[H^{\rm od},T^2]\approx 0$.
The equality is not exact if we use a Hartree-Fock reference state with $N\neq Z$ because the effective one-body part of the Hamiltonian will still break isospin, corresponding to protons and neutrons feeling different mean fields.
Even for a reference with $N=Z$, the equality will also not be exact due to explicit isospin-breaking terms in the Hamiltonian (mainly Coulomb) yielding different single-particle wave functions for protons and neutrons.

The impact of partitioning is illustrated in Fig~\ref{fig:O14N14T2HT}(b) and (c).
In (b), we decouple a $p$-shell valence space (see Fig.~\ref{fig:O14N14valence}), so that $H^{\rm od}$ treats protons and neutrons on equal footing, and then diagonalize in the $p$-shell to obtain $\langle T^2\rangle$ in the $^{14}$O ground state.
If we use a $^{14}$N reference for normal ordering (blue lines), there is no spurious isospin breaking.
If we use an $^{14}$O reference, we find similar isospin breaking to the single-reference case, though the norm of $[H,T]$ is somewhat smaller, especially the two-body part.
Based on this, one might be tempted to conclude that the only relevant choice is the normal ordering reference.
In (c), we decouple a valence space of just neutron $p$-shell orbits.
Here $H^{\rm od}$ treats protons and neutrons unequally, leading to $[\eta,T^2]\neq 0$.
In this case, even with a symmetric $^{14}$N reference we observe substantial symmetry breaking.
Evidently, avoiding spurious isospin breaking requires both a symmetric reference, and a symmetric partitioning.

Below, we identify two additional sources of isospin breaking in the generator; however the impact of these sources is negligible compared with the magnitude of the explicit isospin breaking.

\subsubsection{Energy denominators\label{subsubsec:gen/denom}}
The White, arctangent and imaginary-time generators all feature energy denominators \eqref{eq:DeltaEN1} defined in the introduction.
In our angular momentum coupled implementation, we take the monopole part of the two-body Hamiltonian~\cite{Hergert2016} 
\begin{equation}
    \label{eq:monopole}
    \Gamma_{pqpq}\to \bar{\Gamma}_{pqpq}\equiv \frac{\sum_J (2J+1)\Gamma_{pqpq}^J}{\sum_J (2J+1)},
\end{equation}
when evaluating the denominators in the Epstein-Nesbet partitioning.
This choice breaks isospin symmetry, as can be seen by considering the denominators for $\eta_{abij}^J$ with $a=b$ up to isospin projection.
If $a$ and $b$ are both neutrons, then only even $J$ will contribute to the $\bar{\Gamma}_{abab}$ term, but if $a,b$ are proton-neutron, then odd $J$ will also contribute, leading to a different denominator.
Similar effects modify the other monopole terms.
In principle, one could take more care and maintain isospin in the averaging, but it is much easier to simply use the M{\o}ller-Plesset denominators for which the elements $\Gamma_{pqrs}$ do not appear in the first place.
For the imaginary-time generator, this is even less of an issue in practice because only the sign of the denominator enters the generator, and it is very unlikely that the monopole averaging will change the sign of a denominator.

\subsubsection{Non-linear mixing from the atan generator}
The arctangent generator~\eqref{eq:atan}, along with any generator defined as a formal power series of an argument of the form $H^{\t{od}}/\Delta$, leads to spurious isospin breaking due to non-linear terms in its definition
\begin{equation}
    \label{eq:atan-series}
    \mathrm{atan}(x)=x-\frac{1}{3}x^3+\frac{1}{5}x^5+\ldots \ .
\end{equation}
 Because the expansion is matrix-element-wise, rather than a definition in terms of powers of operators, the non-linear terms spoil the linear relationship between the coupled and uncoupled isospin bases.
 For example, if a given $T_z=0$ element is a superposition of $T=0,1$, the cubic term gives
 \begin{align}
    \label{eq:isomix}
     \eta_{pnpn}^3&=\frac{1}{2}(\eta_{T=1}+\eta_{T=0})^3 \\
     &= \eta_{T=1}^3 + \eta_{T=0}^3 + 3\eta_{T=1}^2\eta_{T=0} + 3\eta_{T=1}\eta_{T=0}^2.\nonumber
 \end{align}
 Because of the cross terms like $\eta_{T=1}^2\eta_{T=0}$ we cannot extract the $\eta_{T=1}$ component by the usual linear combination of $pn$ matrix elements, and we have $[\eta,T^2]\neq 0$.
In practice, this is only a minor effect because usually $H^{\rm od}/\Delta \ll 1$, so that the non-linear terms have a small impact.
 
\section{Benchmark calculations of \texorpdfstring{$\delta_C$}{deltaC}\label{sec:deltaC}}

We now turn to calculations of the isospin-breaking correction $\delta_C$.
For an isospin-conserving Hamiltonian, we should have $\delta_C=0$, so any non-zero $\delta_C$ in this case is an indication of spurious isospin breaking.
However, obtaining $\delta_C=0$ in the isospin conserving case is no guarantee of the accuracy of $\delta_C$ once explicit isospin breaking terms are included.
To assess the accuracy in the presence of explicit isospin breaking, we add the Coulomb interaction to the isospin-conserving interaction $V_{\rm iso}$, and employ several different methods and approximation schemes.

In the IMSRG calculations, we utilize the White generator with Moller-Plesset energy denominators, though our conclusions do not depend strongly on these choices.
We decouple a $p$-shell valence space, diagonalize the effective interaction in the valence space to obtain the initial and final states, and take the transition matrix element of the consistently-transformed $T^{\pm}$ operator.
We perform calculations at the IMSRG(2), IMSRG(3N7) and IMSRG(3) levels.
To the extent that the spurious and explicit/genuine symmetry breaking are additive effects, i.e. $\delta_C \approx \delta_C^{\rm sp}+ \delta_C^{\rm gen}$, we can perform two calculations, one with an isospin conserving Hamiltonian and one with isospin-breaking terms included, and take the difference to isolate the explicit isospin breaking correction. 

It is helpful to compare the IMSRG calculations to canonical perturbation theory~\cite{Primas1963,Klein1974}, which essentially amounts to solving for the Magnus operator $\Omega$ order-by-order in perturbation theory, rather than by integrating a flow equation.
Details are provided in appendix~\ref{app:canonicalPT}.
Because canonical perturbation theory is formulated in terms of commutators, we can apply the same (2), (3N7) and (3) truncations to the perturbative expressions.
As in the IMSRG calculation, we decouple a $p$-shell valence space, diagonalize, and take the transition matrix element of the consistently transformed $T^\pm$ operator.
In principle, to maintain strict order-by-order perturbation theory the interaction should only be evaluated perturbatively in the valence space.
However, we observe that the main effect is due to the transformation of the $T^{\pm}$ operator, so for convenience we simply use the wave functions obtained by diagonalizing the interaction up to fourth order.

As discussed in section~\ref{sec:EOM}, the RPA does not introduce spurious isospin breaking.
This is due to the relatively simple way in which correlations are treated, and so it is less clear how well the RPA captures explicit symmetry breaking.
In our RPA calculations, we perform a Hartree-Fock calculation of the $^{14}$O ground state, and treat $^{14}$N as a charge-changing excitation of the form~\cite{Lane1980} $\chi_\nu^{\dagger}=\sum (X_{a_ni_p} a_n^\dagger i_p - Y_{i_na_p} i_n^{\dagger}a_p)$.

Finally, we compare with configuration interaction (CI) calculations performed in a harmonic oscillator basis, truncated by the total number of oscillator quanta $N_{\rm max}$.
This truncation does not introduce spurious isospin breaking since the $T^\pm$ operator does not change the number of oscillator quanta.
In the complete basis limit this corresponds to the exact result for a given single-particle space.
Here we use a series of truncations $N_{\rm max}=2\ldots 10$ to assess convergence.

\begin{figure}
    \centering
    \includegraphics[width=0.9\columnwidth]{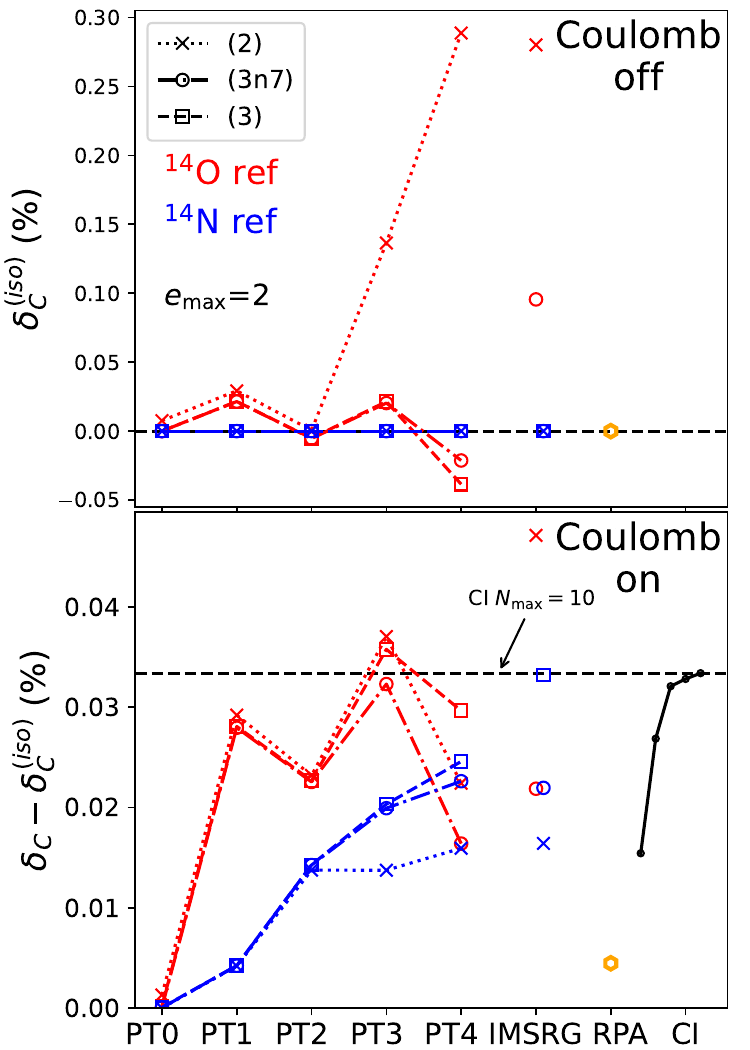}
    \caption{Isospin-breaking correction $\delta_C$ for the $^{14}\mathrm{O}\to^{14}\mathrm{N}$ decay.
    The top panel shows calculations with an isospin-conserving Hamiltonian, while the bottom shows calculations with the Coulomb potential included, with the spurious contribution in the top panel subtracted off. For the CI calculations, the dots represent truncations $N_{\rm max}=2,4,6,8,10$.}
    \label{fig:deltaCperte2}
\end{figure}

All calculations are performed in a single-particle space with $e_{\rm max}=2$.
The results are presented in Fig.~\ref{fig:deltaCperte2}.
In the upper panel, labeled ``Coulomb off'', calculations are performed with the isospin-conserving interaction described in section~\ref{sec:spuriousbreaking}.
Any non-zero $\delta_C$ is spurious.
Next, the Coulomb potential is added to the isospin-conserving Hamiltonian and the calculations are repeated.
The difference between the two calculations is shown in the lower panel.

\begin{figure}
    \centering
    \includegraphics[width=0.9\columnwidth]{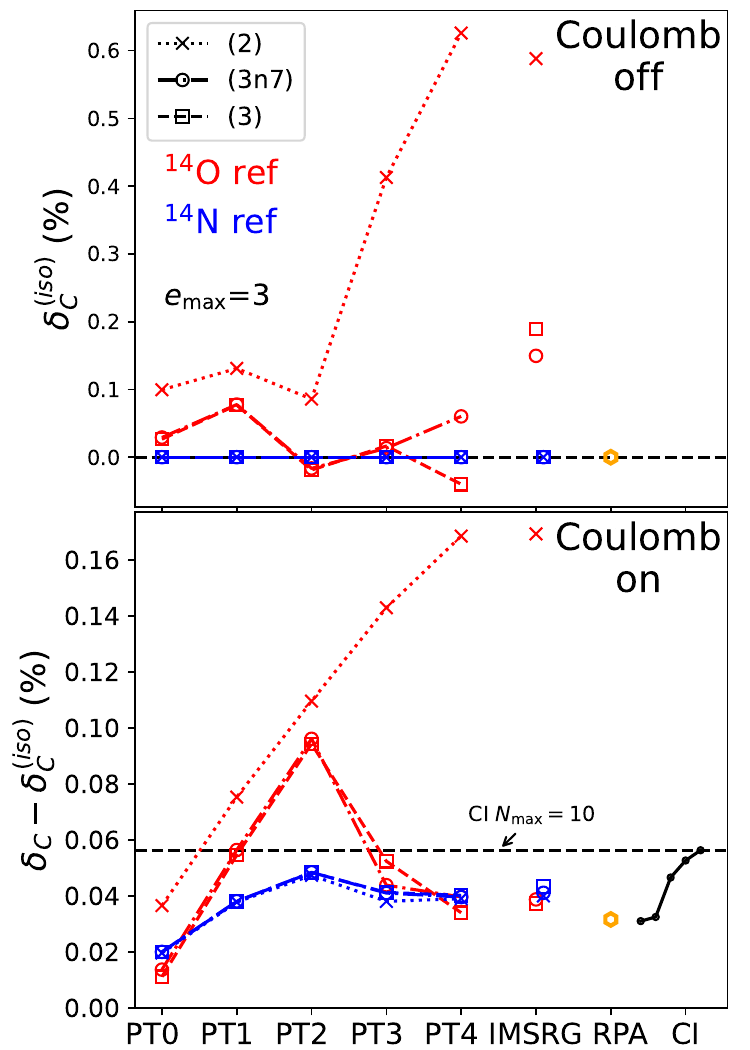}
    \caption{Same as for Fig.~\ref{fig:deltaCperte2}, but for $e_{\rm max}=3$. The black dots correspond to CI calculations at $N_{\rm max}=2,4,6,8,10$.}
    \label{fig:deltaCperte3}
\end{figure}

We observe that, using the $^{14}$N reference, the spurious $\delta_C$ is negligible for all orders of perturbation theory and IMSRG.
With the $^{14}$O reference, on the other hand, we see a substantial spurious contribution, which approaches an order of magnitude larger than the explicit part for the (2) truncation and higher orders of perturbation theory.
Including three-body operators at the (3N7) or (3) level significantly reduces the spurious breaking, but does not eliminate it.
We can also observe oscillations, presumably due to cancellations between different orders in perturbation theory.
With the Coulomb interaction included, we see a significant correction going from IMSRG(3N7) to IMSRG(3), bringing the result into agreement with the CI calculation.
(For the $^{14}$O reference at $e_{\rm max}=2$, the IMSRG(3) evolution failed to converge.)

We also see that, while the RPA does not exhibit spurious isospin breaking, it significantly underpredicts the explicit symmetry breaking.
In fact, this is not entirely surprising, since the RPA approximates excited states as one-particle-one-hole excitations.
At $e_{\rm max}=2$, excitations from the $p$-shell to the $sd$-shell are forbidden by parity, so the RPA can add very little.
We therefore extend the calculation to $e_{\rm max}=3$, shown in Fig.~\ref{fig:deltaCperte3}.
The symmetry-breaking effects of the $^{14}$O reference are qualitatively similar, though quantitatively enhanced.
Notably, the spread in values between different approximations and references is significantly reduced, once the spurious breaking from the $^{14}$O reference has been subtracted.
It appears that the approximate subtraction of spurious contributions, combined with a higher-order truncation scheme, brings the spurious breaking under control.
However, we also find that the resulting values for $\delta_C$ are somewhat smaller than the nearly-converged CI values.
This indicates that, as with the RPA, elimination of the spurious breaking in the IMSRG does not guarantee that the explicit breaking is accurately captured.

\subsection{Perturbative treatment of isospin-breaking terms}
Due to the relative weakness of the isospin-breaking part of the interaction (both strong and Coulomb), it is natural to suspect they could be accurately treated in perturbation theory, and one might hope this could be better controlled~\cite{Seng2026}.
Presumably, the procedure of mixing isospin-projected mean-field states described in~\cite{Satula2009,Satula2012} is essentially perturbative.

The first order contribution to $\delta_C$ vanishes due to the orthogonality of states with different isospin, so the leading correction is second order.
Labeling the exact states by $|\psi_i\rangle$ and $|\psi_f\rangle$, the unperturbed states (with good isospin) as $|i\rangle$, $|f\rangle$, $|x\rangle$, $|y\rangle$, and the isospin-breaking potential $V$, the second-order correction to the matrix element may be written as
\begin{equation}\label{eq:secondorder}
\begin{aligned}
    \langle \psi_f | T^{\pm} | \psi_i\rangle^{(2)} =& \sum_{xy} 
    \frac{\langle f| V | x\rangle} {E_f-E_x}
\langle x|T^{\pm}|y\rangle
    \frac{\langle y| V | i \rangle} {E_i-E_y}\\
     -\frac{1}{2} \langle f|T^{\pm}|i\rangle& \sum_{x} \Bigl( \frac{\langle f|V|x\rangle^2}{(E_f-E_x)^2} + \frac{\langle x|V|i\rangle^2}{(E_i-E_x)^2} \Bigr).
    \end{aligned}
\end{equation}
Explicit evaluation of~\eqref{eq:secondorder} requires accurate knowledge of the excited state spectrum for the isospin-conserving part of the interaction.
Methods like the IMSRG that scale polynomially with system size tend to be more accurate for ground states than for excited states, but it is possible the achieved accuracy would be sufficient.
A challenge specific to the VS-IMSRG is that in its standard formulation one only has access to states within the valence space, while the $|x\rangle,|y\rangle$ in~\eqref{eq:secondorder} can lie outside the valence space.
This motivates the development of an extension to the VS-IMSRG capable of capturing excited states beyond the valence space.
Work in this direction is ongoing and will be presented elsewhere~\cite{SunInPrep}.
In any case, in order to faithfully generate states with good isospin one should take steps to minimize spurious symmetry breaking.

\section{Conclusion}
We have used a toy model to investigate how truncations made in the IMSRG lead to spurious breaking of isospin symmetry, with an eye toward accurate calculations of the correction $\delta_C$ for superallowed $\beta$ decay.
The primary sources of spurious symmetry breaking were found to be the use of a reference $|\Phi_0\rangle$ with $N\neq Z$, and a partitioning $H=H^{\rm d}+H^{\rm od}$ that treats protons and neutrons differently.
Other factors were identified which produced only comparably small effects. 
The essential mechanism for all of these sources of symmetry breaking is the rearrangement of terms which should cancel into expressions involving different intermediate particle ranks.

We demonstrated that results obtained using a reference with $N\neq Z$ could be improved by subtracting off the purely spurious $\delta_C$ correction obtained with an isospin conserving Hamiltonian.
With this adjustment, the agreement between results obtained with different references is significantly enhanced, with systematic improvement as the particle-rank truncation is relaxed.
Unfortunately, large-scale calculations employing the IMSRG(3) or even IMSRG(3N7) approximations are not currently feasible.
An approximate treatment, referred to as IMSRG(3f2) is feasible, and provides a promising path forward.
Implementation and application of the IMSRG(3f2) approximation to the $\delta_C$ correction will be discussed elsewhere.

\acknowledgements{We thank Zhonghao Sun for providing the results of large-$N_{\rm max}$ CI calculations, as well as Bingcheng He, Heiko Hergert, Gaute Hagen, and Thomas Papenbrock for helpful discussions. 
The IMSRG code used in this work~\cite{IMSRG++} utilizes the Armadillo C++ library~\cite{Sanderson2016,Armadillo}. 
SRS acknowledges support from the NSF under Grant No. PHY-2340834, the NSF FRHTP PHY-2402275, and the DOE Topical Collaboration “Nuclear Theory for New Physics,” Award No. DE-SC0023663. AF was supported by the Naughton Fellowship.}

\appendix

\section{IMSRG Generators}

The \textit{White generator}~\cite{White2002} is defined as
\begin{align}\label{eq:whitefull}
    \eta^{\rm Wh}(s)&\equiv\frac{H^\t{od}(s)}{\Delta(s)} \\ 
    &=\sum_{ai}\frac{f_{ai}}{\Delta_{ai}}\{\hat{a}^\dagger \hat{i}\}+\frac{1}{4}\sum_{abij}\frac{\Gamma_{abij}}{\Delta_{abij}}\{\hat{a}^\dagger\hat{b}^\dagger\hat{j}\hat{i}\}
    +\ldots\ .\notag
\end{align}
The \textit{arctangent generator}~\cite{White2002} is 
\begin{align}
\label{eq:atanfull}
    \eta^{\rm atan}(s)&\equiv\frac{1}{2}\mathrm{atan}(2\frac{H^\t{od}(s)}{\Delta})\\
    &=\frac{1}{2}\sum_{ai}\mathrm{atan}(2\frac{f_{ai}}{\Delta_{ai}})\{\hat{a}^\dagger \hat{i}\} \notag \\
    &\qquad+\frac{1}{8}\sum_{ai}\mathrm{atan}(2\frac{\Gamma_{abij}}{\Delta_{abij}})\{\hat{a}^\dagger\hat{b}^\dagger \hat{j}\hat{i}\}+\ldots \notag
\end{align}
while the \textit{imaginary time generator}~\cite{Hergert2016PhysRep} is 
\begin{align}\label{eq:ITfull}
    \eta^{\rm IT}(s)&\equiv \mathrm{sgn}{\Delta} H^\t{od}(s)\\
    &=\sum_{ai}\mathrm{sgn}{\Delta_{ai}}f_{ai}\{\hat{a}^\dagger \hat{i}\} \notag \\
    &\qquad +\frac{1}{4}\sum_{abij}\mathrm{sgn}{\Delta_{abij}}\Gamma_{abij}\{\hat{a}^\dagger\hat{b}^\dagger\hat{j}\hat{i}\}+\ldots\ .\notag
\end{align}

\section{Explicit expressions for Goldstone diagrams\label{app:diagrams}}
Here we provide the expressions corresponding to the Goldstone diagrams shown in Fig.~\ref{fig:CommHT1b}.
\begin{align}
    ({\rm a}) &= \langle a_n i_n | V| a_n i_n \rangle\; \langle a_n | T^\pm | a_p \rangle\; n_{i_n} (1-n_{a_n}) \\
    ({\rm b}) &= \langle a_n i_p | V| a_n i_p \rangle\; \langle a_n | T^\pm | a_p \rangle\; n_{i_p} (1-n_{a_n}) \\
    ({\rm c}) &= \langle a_p i_n | V| a_p i_n \rangle\; \langle a_n | T^\pm | a_p \rangle\; n_{i_n} (1-n_{a_p}) \\
    ({\rm d}) &= \langle a_p i_p | V| a_p i_p \rangle\; \langle a_n | T^\pm | a_p \rangle\; n_{i_p} (1-n_{a_p}) \\
    ({\rm e}) &= \langle a_n i_p | V| a_p i_n \rangle\; \langle i_n | T^\pm | i_p \rangle\; n_{i_p} (1-n_{i_n}) \\
    ({\rm f}) &= \langle a_n i_p | V| a_p i_n  \rangle\; \langle i_n | T^\pm | i_p \rangle\; n_{i_n} (1-n_{i_p})
\end{align}
In the above expressions, $n_x=1$ if orbit $x$ is occupied and $n_x=0$ if it is unoccupied.
For an ensemble reference, we can have $0\leq n_x \leq 1$.
The two-body matrix elements of $V$ are antisymmetrized.

\section{Canonical Perturbation Theory\label{app:canonicalPT}}

In canonical perturbation theory, we partition the Hamiltonian $H=H_0+V$, where $H_0$ contains the diagonal one-body part of $H$, and $V$ is the rest.
With this choice we can invert an equation involving a commutator of the form
\begin{equation}
    [A,H_0] = B.
\end{equation}
Taking a matrix element in an arbitrary Fock-state, we have
\begin{equation}
\begin{aligned}
    \langle i j\ldots | [A,H_0] |k l \ldots \rangle  & =\\ \langle i j \ldots|A|k l \ldots \rangle& (f_{kk}+f_{ll}+\ldots - f_{ii}-f_{jj}-\ldots  ) 
    \end{aligned}
\end{equation}
by which we can obtain the matrix elements of $A$ in terms of $B$:
\begin{equation}
        \langle i j\ldots | A |k l \ldots \rangle   =\frac{ -\langle i j \ldots|B|k l \ldots \rangle}{ f_{ii}+f_{jj}+\ldots - f_{kk}-f_{ll}-\ldots}  
\end{equation}
or in a condensed notation $A=-B/\Delta$.
We now transform $\tilde{H}=e^{\Omega}He^{-\Omega}$, and enforce $\tilde{H}^{\rm od}=0$ order by order in $V$.
(Note the partitioning $H=H^{\rm d}+H^{\rm od}$ is distinct from the partitioning $H=H_0+V$).
This procedure leaves the diagonal part of $\Omega$ unspecified, and we make the convenient choice $\Omega^{\rm d}=0$.
Through fourth order, the expressions for $H$ are
\begin{align}
H^{(0)} &= H_0 \\  
H^{(1)} &= [\Omega^{(1)},H_0] + V  \\
H^{(2)} &= [\Omega^{(2)},H_0] + [\Omega^{(1)},V] + \tfrac{1}{2}[\Omega^{(1)},[\Omega^{(1)},H_0]]
\end{align}
\begin{align}
H^{(3)} &= [\Omega^{(3)},H_0] + [\Omega^{(2)},V] + \tfrac{1}{2}\Bigl( [\Omega^{(1)},[\Omega^{(1)},V]] \nonumber \\
&~~+ [\Omega^{(2)},[\Omega^{(1)},H_0]] + [\Omega^{(1)},[\Omega^{(2)},H_0]] \Bigr) \nonumber \\
&~~+\tfrac{1}{6} [\Omega^{(1)},[\Omega^{(1)},[\Omega^{(1)},H_0]]]
\end{align}
\begin{align}
H^{(4)} &= [\Omega^{(4)},H_0] + [\Omega^{(3)},V] + \tfrac{1}{2}\Bigl([\Omega^{(2)},[\Omega^{(1)},V]]  \nonumber \\
&~~+ [\Omega^{(1)},[\Omega^{(2)},V]] + [\Omega^{(2)},[\Omega^{(2)},H_0]] \nonumber \\
&~~+ [\Omega^{(3)},[\Omega^{(1)},H_0]] + [\Omega^{(1)},[\Omega^{(3)},H_0]] \Bigr) \nonumber \\
&~~+ \tfrac{1}{6}\Bigl(
     [\Omega^{(1)},[\Omega^{(1)},[\Omega^{(1)},V]]]  
   + [\Omega^{(1)},[\Omega^{(1)},[\Omega^{(2)},H_0]]] \nonumber \\
&~~+ [\Omega^{(1)},[\Omega^{(2)},[\Omega^{(1)},H_0]]]
   + [\Omega^{(2)},[\Omega^{(1)},[\Omega^{(1)},H_0]]] \Bigr) \nonumber \\
&~~+\tfrac{1}{24} [\Omega^{(1)},[\Omega^{(1)},[\Omega^{(1)},[\Omega^{(1)},H_0]]]].
\end{align}
The expression for $\Omega^{(n)}$ can be obtained from the $H^{(n)}$ equation by taking the off-diagonal piece of both sides and setting the left hand side to zero, yielding an equation for $\Omega^{(n)}$ that only depends on lower orders.
Through second order we have
\begin{align}
\Omega^{(0)} &= 0 \\
\Omega^{(1)} &= V^{\rm od}/\Delta \\
\Omega^{(2)} &= \Bigl( [\Omega^{(1)},V]^{\rm od} + \tfrac{1}{2}[\Omega^{(1)},[\Omega^{(1)},H_0]]^{\rm od} \Bigr) \bigr/ \Delta.
\end{align}
Finally, we may consistently transform another operator $\mathcal{O}$ as $\tilde{\mathcal{O}}=e^{\Omega}\mathcal{O}e^{-\Omega}$.
Through second order this is
\begin{align}
    \tilde{\mathcal{O}}^{(0)} &= \mathcal{O} \\
    \tilde{\mathcal{O}}^{(1)} &= [\Omega^{(1)},\mathcal{O}] \\
    \tilde{\mathcal{O}}^{(2)} &= [\Omega^{(2)},\mathcal{O}] + \tfrac{1}{2}[\Omega^{(1)},[\Omega^{(1)},\mathcal{O}]]. 
\end{align}

\bibliography{references}

@article{Cabibbo:1963yz,
    author = "Cabibbo, Nicola",
    title = "{Unitary Symmetry and Leptonic Decays}",
    doi = "10.1103/PhysRevLett.10.531",
    journal = "Phys. Rev. Lett.",
    volume = "10",
    pages = "531--533",
    year = "1963"
}

@article{Kobayashi:1973fv,
    author = "Kobayashi, Makoto and Maskawa, Toshihide",
    title = "{CP Violation in the Renormalizable Theory of Weak Interaction}",
    reportNumber = "KUNS-242",
    doi = "10.1143/PTP.49.652",
    journal = "Prog. Theor. Phys.",
    volume = "49",
    pages = "652--657",
    year = "1973"
}

@book{ShavittBartlett,
author = {Shavitt, Isaiah and Bartlett, Rodney},
year = {2009},
month = {01},
pages = {},
title = {Many-Body Methods in Chemistry and Physics: MBPT and Coupled-Cluster Theory},
isbn = {9780521818322},
journal = {Many - Body Methods in Chemistry and Physics: MBPT and Coupled-Cluster Theory},
doi = {10.1017/CBO9780511596834},
publisher = {Cambridge University Press}
}

@article{Magnus,
  title = {Magnus expansion and in-medium similarity renormalization group},
  author = {Morris, T. D. and Parzuchowski, N. M. and Bogner, S. K.},
  journal = {Phys. Rev. C},
  volume = {92},
  issue = {3},
  pages = {034331},
  numpages = {12},
  year = {2015},
  month = {Sep},
  publisher = {American Physical Society},
  doi = {10.1103/PhysRevC.92.034331},
  url = {https://link.aps.org/doi/10.1103/PhysRevC.92.034331}
}

@article{TownerHardy2020,
  title = {{Superallowed ${0}^{+}\ensuremath{\rightarrow}{0}^{+}$ nuclear $\ensuremath{\beta}$ decays: 2020 critical survey, with implications for ${V}_{\mathit{ud}}$ and CKM unitarity}},
  author = {Hardy, J. C. and Towner, I. S.},
  journal = {Phys. Rev. C},
  volume = {102},
  issue = {4},
  pages = {045501},
  numpages = {29},
  year = {2020},
  month = {Oct},
  publisher = {American Physical Society},
  doi = {10.1103/PhysRevC.102.045501},
  url = {https://link.aps.org/doi/10.1103/PhysRevC.102.045501}
}

@article{Hergert2016,
	doi = {10.1088/1402-4896/92/2/023002},
	url = {https://doi.org/10.1088%2F1402-4896%2F92%2F2%2F023002},
	year = {2016},
	month = {dec},
	publisher = {{IOP} Publishing},
	volume = {92},
	number = {2},
	pages = {023002},
	author = {H Hergert},
	title = {In-medium similarity renormalization group for closed and open-shell nuclei},
	journal = {Phys. Scr.}
}

@article{Heinz2021,
  title = {In-medium similarity renormalization group with three-body operators},
  author = {Heinz, M. and Tichai, A. and Hoppe, J. and Hebeler, K. and Schwenk, A.},
  journal = {Phys. Rev. C},
  volume = {103},
  issue = {4},
  pages = {044318},
  numpages = {21},
  year = {2021},
  month = {Apr},
  publisher = {American Physical Society},
  doi = {10.1103/PhysRevC.103.044318},
  url = {https://link.aps.org/doi/10.1103/PhysRevC.103.044318}
}

@misc{IMSRG++,
author = {Stroberg, Steven Ragnar},
license = {GPL-2.0-or-later},
month = sep,
title = {{IMSRG++}},
url = {https://github.com/ragnarstroberg/imsrg},
howpublished = {https://github.com/ragnarstroberg/imsrg},
version = {0.1.0},
year = {2018}
}

@article{stroberg2024imsrgflowing3body,
  title = {In-medium similarity renormalization group with flowing 3-body operators, and approximations thereof},
  author = {Stroberg, S. R. and Morris, T. D. and He, B. C.},
  journal = {Phys. Rev. C},
  volume = {110},
  issue = {4},
  pages = {044316},
  numpages = {16},
  year = {2024},
  month = {Oct},
  publisher = {American Physical Society},
  doi = {10.1103/PhysRevC.110.044316},
  url = {https://link.aps.org/doi/10.1103/PhysRevC.110.044316}
}

@article{he2024factorizedapproximationimsrg3,
  title = {{Factorized approximation to the in-medium similarity renormalization group IMSRG(3)}},
  author = {He, B. C. and Stroberg, S. R.},
  journal = {Phys. Rev. C},
  volume = {110},
  issue = {4},
  pages = {044317},
  numpages = {16},
  year = {2024},
  month = {Oct},
  publisher = {American Physical Society},
  doi = {10.1103/PhysRevC.110.044317},
  url = {https://link.aps.org/doi/10.1103/PhysRevC.110.044317}
}

@article{Yuan2022,
  title = {Deformed in-medium similarity renormalization group},
  author = {Yuan, Q. and Fan, S. Q. and Hu, B. S. and Li, J. G. and Zhang, S. and Wang, S. M. and Sun, Z. H. and Ma, Y. Z. and Xu, F. R.},
  journal = {Phys. Rev. C},
  volume = {105},
  issue = {6},
  pages = {L061303},
  numpages = {7},
  year = {2022},
  month = {Jun},
  publisher = {American Physical Society},
  doi = {10.1103/PhysRevC.105.L061303},
  url = {https://link.aps.org/doi/10.1103/PhysRevC.105.L061303}
}

@article{White2002,
    author = {White, Steven R.},
    title = {Numerical canonical transformation approach to quantum many-body problems},
    journal = {J. Chem. Phys.},
    volume = {117},
    number = {16},
    pages = {7472-7482},
    year = {2002},
    month = {10},
    issn = {0021-9606},
    doi = {10.1063/1.1508370},
    url = {https://doi.org/10.1063/1.1508370},
}

@article{Hergert2016PhysRep,
title = {The In-Medium Similarity Renormalization Group: A novel ab initio method for nuclei},
journal = {Phys. Rep.},
volume = {621},
pages = {165-222},
year = {2016},
issn = {0370-1573},
doi = {https://doi.org/10.1016/j.physrep.2015.12.007},
url = {https://www.sciencedirect.com/science/article/pii/S0370157315005414},
author = {H. Hergert and S.K. Bogner and T.D. Morris and A. Schwenk and K. Tsukiyama},
}

@article{Stroberg2019,
  title={Nonempirical interactions for the nuclear shell model: an update},
  author={Stroberg, S Ragnar and Hergert, Heiko and Bogner, Scott K and Holt, Jason D},
  journal={Ann. Rev. Nucl. Part. Sci.},
  volume={69},
  number={1},
  pages={307--362},
  year={2019},
  doi={10.1146/annurev-nucl-101917-021120},
  publisher={Annual Reviews}
}

@article{Stroberg2017valence,
  title = {Nucleus-Dependent Valence-Space Approach to Nuclear Structure},
  author = {Stroberg, S. R. and Calci, A. and Hergert, H. and Holt, J. D. and Bogner, S. K. and Roth, R. and Schwenk, A.},
  journal = {Phys. Rev. Lett.},
  volume = {118},
  issue = {3},
  pages = {032502},
  numpages = {6},
  year = {2017},
  month = {Jan},
  publisher = {American Physical Society},
  doi = {10.1103/PhysRevLett.118.032502},
  url = {https://link.aps.org/doi/10.1103/PhysRevLett.118.032502}
}

@article{Cirigliano_2023,
   title={{Scrutinizing CKM unitarity with a new measurement of the K3/K2 branching fraction}},
   volume={838},
   ISSN={0370-2693},
   url={http://dx.doi.org/10.1016/j.physletb.2023.137748},
   DOI={10.1016/j.physletb.2023.137748},
   journal={Phys. Lett. B},
   publisher={Elsevier BV},
   author={Cirigliano, Vincenzo and Crivellin, Andreas and Hoferichter, Martin and Moulson, Matthew},
   year={2023},
   month=mar, pages={137748} }

@Unpublished{chakraborty2024workinggroup1summary,
      title={{Working group 1 summary: $V_{ud}$, $V_{us}$, $V_{cd}$, $V_{cs}$ and semileptonic/leptonic $D$ decays}}, 
      author={Bipasha Chakraborty and Alex Gilman and Martin Hoferichter and Michal Koval},
      year={2024},
      eprint={2406.13703},
      archivePrefix={arXiv},
      primaryClass={hep-ph},
      url={https://arxiv.org/abs/2406.13703}, 
}

@article{Jiang2020deltaGO,
  title = {{Accurate bulk properties of nuclei from $A=2$ to $\ensuremath{\infty}$ from potentials with $\mathrm{\ensuremath{\Delta}}$ isobars}},
  author = {Jiang, W. G. and Ekstr\"om, A. and Forss\'en, C. and Hagen, G. and Jansen, G. R. and Papenbrock, T.},
  journal = {Phys. Rev. C},
  volume = {102},
  issue = {5},
  pages = {054301},
  numpages = {8},
  year = {2020},
  month = {Nov},
  publisher = {American Physical Society},
  doi = {10.1103/PhysRevC.102.054301},
  url = {https://link.aps.org/doi/10.1103/PhysRevC.102.054301}
}

@article{Brink1970,
  title={{Isospin mixing of Hartree-Fock solutions}},
  author={Brink, DM and Svenne, JP},
  journal={Nuc. Phys. A},
  volume={154},
  number={3},
  pages={449--457},
  year={1970},
  doi={10.1016/0375-9474(70)90117-X},
  publisher={Elsevier}
}

@article{barrett2013,
  title={Ab initio no core shell model},
  author={Barrett, Bruce R and Navr{\'a}til, Petr and Vary, James P},
  journal={Prog. Part. Nucl. Phys.},
  volume={69},
  pages={131--181},
  year={2013},
  doi={10.1016/j.ppnp.2012.10.003},
  publisher={Elsevier}
}

@article{Caurier2002,
  title = {Ab initio shell model for $A=10$ nuclei},
  author = {Caurier, E. and Navr\'atil, P. and Ormand, W. E. and Vary, J. P.},
  journal = {Phys. Rev. C},
  volume = {66},
  issue = {2},
  pages = {024314},
  numpages = {14},
  year = {2002},
  month = {Aug},
  publisher = {American Physical Society},
  doi = {10.1103/PhysRevC.66.024314},
  url = {https://link.aps.org/doi/10.1103/PhysRevC.66.024314}
}

@article{Kutzelnigg1997,
    author = {Kutzelnigg, Werner and Mukherjee, Debashis},
    title = {{Normal order and extended Wick theorem for a multiconfiguration reference wave function}},
    journal = {J. Chem. Phys.},
    volume = {107},
    number = {2},
    pages = {432-449},
    year = {1997},
    month = {07},
    issn = {0021-9606},
    doi = {10.1063/1.474405},
    url = {https://doi.org/10.1063/1.474405},
}

@article{Baym1961,
  title = {Conservation Laws and Correlation Functions},
  author = {Baym, Gordon and Kadanoff, Leo P.},
  journal = {Phys. Rev.},
  volume = {124},
  issue = {2},
  pages = {287--299},
  numpages = {0},
  year = {1961},
  month = {Oct},
  publisher = {American Physical Society},
  doi = {10.1103/PhysRev.124.287},
  url = {https://link.aps.org/doi/10.1103/PhysRev.124.287}
}

@Unpublished{Seng2026,
      title={A comprehensive theory framework for perturbative calculations of $\delta_C$ in superallowed beta decays}, 
      author={Chien-Yeah Seng},
      year={2026},
      eprint={2609.20448},
      archivePrefix={arXiv},
      primaryClass={nucl-th},
      url={https://arxiv.org/abs/2609.20448}, 
}

@book{RingSchuck1980,
  title = {The Nuclear Many-Body Proglem},
  author = {Ring, P. and Schuck, P.},
  year = {1980},
  publisher = {Springer-Verlag}
}

@article{Sheikh2021,
doi = {10.1088/1361-6471/ac288a},
url = {https://doi.org/10.1088/1361-6471/ac288a},
year = {2021},
month = {nov},
publisher = {IOP Publishing},
volume = {48},
number = {12},
pages = {123001},
author = {Sheikh, J A and Dobaczewski, J and Ring, P and Robledo, L M and Yannouleas, C},
title = {Symmetry restoration in mean-field approaches},
journal = {J. Phys. G},
}

@article{Heinz2025,
  title = {Improved structure of calcium isotopes from ab initio calculations},
  author = {Heinz, M. and Miyagi, T. and Stroberg, S. R. and Tichai, A. and Hebeler, K. and Schwenk, A.},
  journal = {Phys. Rev. C},
  volume = {111},
  issue = {3},
  pages = {034311},
  numpages = {14},
  year = {2025},
  month = {Mar},
  publisher = {American Physical Society},
  doi = {10.1103/PhysRevC.111.034311},
  url = {https://link.aps.org/doi/10.1103/PhysRevC.111.034311}
}

@article{Stroberg2021,
  title={Beta decay in medium-mass nuclei with the in-medium similarity renormalization group},
  author={Stroberg, Steven Ragnar},
  journal={Particles},
  volume={4},
  number={4},
  pages={521--535},
  year={2021},
  doi={10.3390/particles4040038},
  publisher={MDPI}
}

@article{Satula2009,
  title = {Isospin Mixing in Nuclei within the Nuclear Density Functional Theory},
  author = {Satu{\l{}}a, W. and Dobaczewski, J. and Nazarewicz, W. and Rafalski, M.},
  journal = {Phys. Rev. Lett.},
  volume = {103},
  issue = {1},
  pages = {012502},
  numpages = {4},
  year = {2009},
  month = {Jul},
  publisher = {American Physical Society},
  doi = {10.1103/PhysRevLett.103.012502},
  url = {https://link.aps.org/doi/10.1103/PhysRevLett.103.012502}
}

@article{Satula2012,
  title = {Isospin-breaking corrections to superallowed Fermi $\ensuremath{\beta}$ decay in isospin- and angular-momentum-projected nuclear density functional theory},
  author = {Satu{\l{}}a, W. and Dobaczewski, J. and Nazarewicz, W. and Werner, T. R.},
  journal = {Phys. Rev. C},
  volume = {86},
  issue = {5},
  pages = {054316},
  numpages = {13},
  year = {2012},
  month = {Nov},
  publisher = {American Physical Society},
  doi = {10.1103/PhysRevC.86.054316},
  url = {https://link.aps.org/doi/10.1103/PhysRevC.86.054316}
}

@article{Rowe1968,
  title = {Equations-of-Motion Method and the Extended Shell Model},
  author = {Rowe, D. J.},
  journal = {Rev. Mod. Phys.},
  volume = {40},
  issue = {1},
  pages = {153--166},
  numpages = {0},
  year = {1968},
  month = {Jan},
  publisher = {American Physical Society},
  doi = {10.1103/RevModPhys.40.153},
  url = {https://link.aps.org/doi/10.1103/RevModPhys.40.153}
}

@book{Rowe2010,
   title = {Nuclear Collective Motion: Models and Theory},
   author = {Rowe, David J},
   publisher = {World Scientific},
   year = {2010}
}

@article{Thouless1961,
title = {Vibrational states of nuclei in the random phase approximation},
journal = {Nucl. Phys.},
volume = {22},
number = {1},
pages = {78-95},
year = {1961},
issn = {0029-5582},
doi = {https://doi.org/10.1016/0029-5582(61)90364-9},
url = {https://www.sciencedirect.com/science/article/pii/0029558261903649},
author = {D.J. Thouless},
}

@misc{SunInPrep,
   author = {Sun, Zhonghao and Stroberg, S. R. },
   howpublished = {In preparation}
}

@article{VanKolck1996,
title = {Phenomenological aspects of isospin violation in the nuclear force},
journal = {Phys. Lett. B},
volume = {371},
number = {3},
pages = {169-174},
year = {1996},
issn = {0370-2693},
doi = {https://doi.org/10.1016/0370-2693(96)00009-3},
url = {https://www.sciencedirect.com/science/article/pii/0370269396000093},
author = {U. {van Kolck} and J.L. Friar and T. Goldman},
}

@article{Piarulli2026,
  title = {{Quantum Monte Carlo calculation of ${\ensuremath{\delta}}_{C}$ in the superallowed $\ensuremath{\beta}$ decay of $^{10}\mathrm{C}$}},
  author = {Piarulli, Maria and Wiringa, R. B. and Lovato, Alessandro and King, Garrett B. and Pastore, Saori},
  journal = {Phys. Rev. C},
  volume = {114},
  issue = {2},
  pages = {024326},
  numpages = {12},
  year = {2026},
  month = {Aug},
  publisher = {American Physical Society},
  doi = {10.1103/5gb3-q3lz},
  url = {https://link.aps.org/doi/10.1103/5gb3-q3lz}
}

@article{Parzuchowski2017,
  title = {Ab initio electromagnetic observables with the in-medium similarity renormalization group},
  author = {Parzuchowski, N. M. and Stroberg, S. R. and Navr{\'a}til, P. and Hergert, H. and Bogner, S. K.},
  journal = {Phys. Rev. C},
  volume = {96},
  issue = {3},
  pages = {034324},
  numpages = {17},
  year = {2017},
  month = {Sep},
  publisher = {American Physical Society},
  doi = {10.1103/PhysRevC.96.034324},
  url = {https://link.aps.org/doi/10.1103/PhysRevC.96.034324}
}

@inproceedings{Armadillo,
   title={{Armadillo: an Efficient Framework for Numerical Linear Algebra}},
   url={http://dx.doi.org/10.1109/ICCAE64891.2025.10980539},
   DOI={10.1109/iccae64891.2025.10980539},
   booktitle={2025 17th International Conference on Computer and Automation Engineering (ICCAE)},
   publisher={IEEE},
   author={Sanderson, Conrad and Curtin, Ryan},
   year={2025},
   month=Mar, pages={303–307} }

@article{Sanderson2016,
  doi = {10.21105/joss.00026},
  url = {https://doi.org/10.21105/joss.00026},
  year = {2016},
  publisher = {The Open Journal},
  volume = {1},
  number = {2},
  pages = {26},
  author = {Conrad Sanderson and Ryan Curtin},
  title = {{Armadillo: a template-based C++ library for linear algebra}},
  journal = {J. Open Source Softw.}
}

@article{Soma2020,
  title={Self-consistent Green's function theory for atomic nuclei},
  author={Soma, Vittorio},
  journal={Front. Phys.},
  volume={8},
  pages={340},
  year={2020},
  doi={10.3389/fphy.2020.00340},
  publisher={Frontiers Media SA}
}

@article{Barbieri2016,
    author = "Barbieri, Carlo and Carbone, Arianna",
    title = "{Self-consistent Green's function approaches}",
    doi = "10.1007/978-3-319-53336-0_11",
    journal = "Lect. Notes Phys.",
    volume = "936",
    pages = "571--644",
    year = "2017"
}

@article{Dickhoff2004,
    author = "Dickhoff, W. H. and Barbieri, C.",
    title = "{{Self-consistent Green's function method for nuclei and nuclear matter}}",
    reportNumber = "TRI-PP-03-40",
    doi = "10.1016/j.ppnp.2004.02.038",
    journal = "Prog. Part. Nucl. Phys.",
    volume = "52",
    pages = "377--496",
    year = "2004"
}

@book{DickhoffMBT,
    author = {Dickhoff, Willem H and Van Neck, Dimitri},
    title = {Many-Body Theory Exposed!},
    subtitle = {Propagator description of quantum mechanics in many-body systems},
    publisher = {World Scientific},
    year = {2005}
}

@article{Ekstrom2014,
  title = {{Effects of Three-Nucleon Forces and Two-Body Currents on Gamow-Teller Strengths}},
  author = {Ekstr{\"o}m, A. and Jansen, G. R. and Wendt, K. A. and Hagen, G. and Papenbrock, T. and Bacca, S. and Carlsson, B. and Gazit, D.},
  journal = {Phys. Rev. Lett.},
  volume = {113},
  issue = {26},
  pages = {262504},
  numpages = {6},
  year = {2014},
  month = {Dec},
  publisher = {American Physical Society},
  doi = {10.1103/PhysRevLett.113.262504},
  url = {https://link.aps.org/doi/10.1103/PhysRevLett.113.262504}
}

@article{Stanton1993,
    author = {Stanton, John F. and Bartlett, Rodney J.},
    title = {{The equation of motion coupled‐cluster method. A systematic biorthogonal approach to molecular excitation energies, transition probabilities, and excited state properties}},
    journal = {J. Chem. Phys.},
    volume = {98},
    number = {9},
    pages = {7029-7039},
    year = {1993},
    month = {05},
    issn = {0021-9606},
    doi = {10.1063/1.464746},
    url = {https://doi.org/10.1063/1.464746},
}

@article{Hagen2014,
doi = {10.1088/0034-4885/77/9/096302},
url = {https://doi.org/10.1088/0034-4885/77/9/096302},
year = {2014},
month = {sep},
publisher = {IOP Publishing},
volume = {77},
number = {9},
pages = {096302},
author = {Hagen, G and Papenbrock, T and Hjorth-Jensen, M and Dean, D J},
title = {Coupled-cluster computations of atomic nuclei},
journal = {Rep. Prog. Phys.},
}

@article{Sawicki1962,
  title = {Higher Random Phase Approximation and Energy Spectra of Spherical Nuclei},
  author = {Sawicki, J.},
  journal = {Phys. Rev.},
  volume = {126},
  issue = {6},
  pages = {2231--2238},
  numpages = {0},
  year = {1962},
  month = {Jun},
  publisher = {American Physical Society},
  doi = {10.1103/PhysRev.126.2231},
  url = {https://link.aps.org/doi/10.1103/PhysRev.126.2231}
}

@article{Qiu2017,
    author = {Qiu, Yiheng and Henderson, Thomas M. and Zhao, Jinmo and Scuseria, Gustavo E.},
    title = {Projected coupled cluster theory},
    journal = {J. Chem. Phys.},
    volume = {147},
    number = {6},
    pages = {064111},
    year = {2017},
    month = {08},
    issn = {0021-9606},
    doi = {10.1063/1.4991020},
    url = {https://doi.org/10.1063/1.4991020},
}

@article{Hagen2022,
  title = {{Angular-momentum projection in coupled-cluster theory: Structure of $^{34}\mathrm{Mg}$}},
  author = {Hagen, G. and Novario, S. J. and Sun, Z. H. and Papenbrock, T. and Jansen, G. R. and Lietz, J. G. and Duguet, T. and Tichai, A.},
  journal = {Phys. Rev. C},
  volume = {105},
  issue = {6},
  pages = {064311},
  numpages = {23},
  year = {2022},
  month = {Jun},
  publisher = {American Physical Society},
  doi = {10.1103/PhysRevC.105.064311},
  url = {https://link.aps.org/doi/10.1103/PhysRevC.105.064311}
}

@article{Duguet2015,
doi = {10.1088/0954-3899/42/2/025107},
url = {https://doi.org/10.1088/0954-3899/42/2/025107},
year = {2014},
month = {dec},
publisher = {IOP Publishing},
volume = {42},
number = {2},
pages = {025107},
author = {Duguet, T},
title = {{Symmetry broken and restored coupled-cluster theory: I. Rotational symmetry and angular momentum}},
journal = {J. Phys. G},
}

@article{Parzuchowski2017x,
  title = {Ab initio excited states from the in-medium similarity renormalization group},
  author = {Parzuchowski, N. M. and Morris, T. D. and Bogner, S. K.},
  journal = {Phys. Rev. C},
  volume = {95},
  issue = {4},
  pages = {044304},
  numpages = {15},
  year = {2017},
  month = {Apr},
  publisher = {American Physical Society},
  doi = {10.1103/PhysRevC.95.044304},
  url = {https://link.aps.org/doi/10.1103/PhysRevC.95.044304}
}

@article{Simkovic2013,
  title = {$0\ensuremath{\nu}\ensuremath{\beta}\ensuremath{\beta}$ and $2\ensuremath{\nu}\ensuremath{\beta}\ensuremath{\beta}$ nuclear matrix elements, quasiparticle random-phase approximation, and isospin symmetry restoration},
  author = { \v{S}imkovic, Fedor and Rodin, Vadim and Faessler, Amand and Vogel, Petr},
  journal = {Phys. Rev. C},
  volume = {87},
  issue = {4},
  pages = {045501},
  numpages = {9},
  year = {2013},
  month = {Apr},
  publisher = {American Physical Society},
  doi = {10.1103/PhysRevC.87.045501},
  url = {https://link.aps.org/doi/10.1103/PhysRevC.87.045501}
}

@article{Engelbrecht1970,
  title = {Isospin Polarization in the Nuclear Many-Body Problem},
  author = {Engelbrecht, C. A. and Lemmer, R. H.},
  journal = {Phys. Rev. Lett.},
  volume = {24},
  issue = {11},
  pages = {607--611},
  numpages = {0},
  year = {1970},
  month = {Mar},
  publisher = {American Physical Society},
  doi = {10.1103/PhysRevLett.24.607},
  url = {https://link.aps.org/doi/10.1103/PhysRevLett.24.607}
}

@article{Lane1980,
title = {{The Random Phase Approximation: Its role in restoring symmetries lacking in the Hartree-Fock approximation}},
journal = {Annals of Physics},
volume = {129},
number = {2},
pages = {273-302},
year = {1980},
issn = {0003-4916},
doi = {10.1016/0003-4916(80)90389-9},
url = {https://www.sciencedirect.com/science/article/pii/0003491680903899},
author = {A.M Lane and J Martorell},
}

@article{Jansen2013,
  title = {Spherical coupled-cluster theory for open-shell nuclei},
  author = {Jansen, G. R.},
  journal = {Phys. Rev. C},
  volume = {88},
  issue = {2},
  pages = {024305},
  numpages = {20},
  year = {2013},
  month = {Aug},
  publisher = {American Physical Society},
  doi = {10.1103/PhysRevC.88.024305},
  url = {https://link.aps.org/doi/10.1103/PhysRevC.88.024305}
}

@article{Hagen2009,
  title = {Solution of the Center-Of-Mass Problem in Nuclear Structure Calculations},
  author = {Hagen, G. and Papenbrock, T. and Dean, D. J.},
  journal = {Phys. Rev. Lett.},
  volume = {103},
  issue = {6},
  pages = {062503},
  numpages = {4},
  year = {2009},
  month = {Aug},
  publisher = {American Physical Society},
  doi = {10.1103/PhysRevLett.103.062503},
  url = {https://link.aps.org/doi/10.1103/PhysRevLett.103.062503}
}

@article{Primas1963,
  title = {Generalized Perturbation Theory in Operator Form},
  author = {Primas, H.},
  journal = {Rev. Mod. Phys.},
  volume = {35},
  issue = {3},
  pages = {710--711},
  numpages = {0},
  year = {1963},
  month = {Jul},
  publisher = {American Physical Society},
  doi = {10.1103/RevModPhys.35.710},
  url = {https://link.aps.org/doi/10.1103/RevModPhys.35.710}
}

@article{Klein1974,
    author = {Klein, D. J.},
    title = {Degenerate perturbation theory},
    journal = {J. Chem. Phys.},
    volume = {61},
    number = {3},
    pages = {786-798},
    year = {1974},
    month = {08},
    issn = {0021-9606},
    doi = {10.1063/1.1682018},
    url = {https://doi.org/10.1063/1.1682018},
}

@Unpublished{Porro2026,
      title={Revisiting the Equation-of-Motion Method: A Universal Framework for Correlated Quantum Systems}, 
      author={Andrea Porro},
      year={2026},
      eprint={2607.29396},
      archivePrefix={arXiv},
      primaryClass={nucl-th},
      url={https://arxiv.org/abs/2607.29396}, 
}

@article{Cheung1980,
title = {Charge symmetry breaking in the neutron-proton system},
journal = {Nuc. Phys. A},
volume = {348},
number = {4},
pages = {365-387},
year = {1980},
issn = {0375-9474},
doi = {https://doi.org/10.1016/0375-9474(80)90261-4},
url = {https://www.sciencedirect.com/science/article/pii/0375947480902614},
author = {C.Y. Cheung and E.M. Henley and G.A. Miller},
}

@article{Bogner2014,
  title={Nonperturbative shell-model interactions from the in-medium similarity renormalization group},
  author={Bogner, SK and Hergert, H and Holt, JD and Schwenk, A and Binder, S and Calci, A and Langhammer, J and Roth, R},
  journal={Phys. Rev. Lett.},
  volume={113},
  number={14},
  pages={142501},
  year={2014},
  publisher={APS}
}

%
%
\end{document}